\documentclass[letterpaper,10pt,conference]{ieeeconf}

\IEEEoverridecommandlockouts    

\usepackage{cite}
\usepackage{amsmath,amssymb,amsfonts}
\usepackage{graphicx}
\usepackage{textcomp}

\usepackage{amssymb} 
\usepackage{comment}
\usepackage{xcolor}
\usepackage[colorlinks=true]{hyperref}
\usepackage{subcaption}

\usepackage{amsthm}
\usepackage{newtxtext,newtxmath}
\usepackage{algorithm}
\usepackage{algpseudocode}
\usepackage{forest}
\usepackage{dsfont}
\usepackage{bbm}
\usepackage{xurl}

\newtheorem{proposition}{Proposition}
\newtheorem{corollary}{Corollary}
\newtheorem{lemma}{Lemma}

\newcounter{onlineappendix}

\title{\LARGE \bf
Personalized and Trust-Aware Health Recommendation Policies\\ for a Construction Workplace
}

\author{Atefeh Mollabagher, Yogesh Gautam, Houtan Jebelli, and Parinaz Naghizadeh%
\thanks{This work is supported in part by the NSF under award \#2411000.}%
\thanks{A. Mollabagher and P. Naghizadeh are with the Electrical and Computer Engineering Department, University of California, San Diego, CA 92162 USA
        {\tt\small \{atefeh,\,parinaz\}@ucsd.edu}.}%
\thanks{Y. Gautam and H. Jebelli are with the Department of Civil and Environmental Engineering, University of Illinois Urbana-Champaign, Urbana, IL
        {\tt\small \{ygautam2,\,hjebelli\}@illinois.edu}.}% 
}

\begin{document}

\maketitle

%%%%%%%%%%%%%%%%%%%%%%%%%%%%%%%%%%%%%%%%%%%%%%%%%%%%%%%%%%%%%%%%%%%%%%%%%%

\begin{abstract}
Construction workers face workplace risks such as fatigue, heat stress, and other physically demanding conditions that can negatively affect their health and safety. Although monitoring these risks is important, timely and personalized health interventions are also needed to help prevent negative impacts on workers' well-being and productivity. To this end, in this paper, we propose a model to capture the interactions between a trust-aware health recommender system and workers who differ in health and trust sensitivity. Specifically, in our proposed dynamic model,  worker health evolves over time, worker trust is affected by both health and recommendation dynamics, and trust in turn affects compliance with future recommendations. Given this model, we characterize the recommender policy, including a health-based recommendation triggering threshold and the recommendation frequency. We do so using both model-based short-horizon control and model-free reinforcement learning. We then investigate how recommendation frequencies are adjusted for different workers to balance their health, productivity, and trust. Our findings provide insight into the design of personalized health recommendation policies in construction workplaces and beyond.
\end{abstract}

%%%%%%%%%%%%%%%%%%%%%%%%%%%%%%%%%%%%%%%%%%%%%%%%%%%%%%%%%%%%%%%%%%%%%%%%%%
 
\section{Introduction}\label{sec:introduction}
The physically demanding nature of construction work, together with exposure to hazards such as fatigue and heat stress, can adversely affect workers' health, attention, and safety \cite{zong2024fatigue, acharya2018assessing, torbat2024heat}. These challenges have led to an increasing interest in monitoring and prevention tools, including wearable sensors \cite{aryal2017monitoring}, heat-stress assessment methods \cite{shakerian2021assessing}, and smart personal protective equipment for construction safety \cite{rasouli2024smart}. While these developments have improved the ability to monitor worker state, many existing systems remain focused on detection or assessment, rather than on deciding how and when such information should be used to generate effective interventions over time. This suggests the need for systems that can use information about the workers' current state to provide timely and personalized interventions to improve health and safety in construction workplaces. Our earlier work \cite{gautam2026memory} has developed such health recommender systems through an agentic AI framework, and demonstrates the potential of memory-augmented agentic AI to deliver timely, personalized, and context-aware guidance.

The effectiveness of such systems depends not only on their ability to detect worker risk, but also on whether workers \emph{trust} the system and are willing to rely on it over time. Recent work on human-AI and human-robot collaboration in construction suggests that trust plays an important role in whether such systems are adopted and used effectively on work sites, and that this trust can be affected by repeated interaction \cite{chang2024toward} and by workers' perceptions of safety and reliability \cite{emaminejad2024assessing}. While these existing works focus on trust in robots and AI-powered collaborative systems in construction workplaces, they suggest a similar effect for our setting of health recommendations. In particular, they suggest that trust should not be treated as static, but rather as an \emph{endogenously-evolving} state that affects how workers respond to the system. A recommendation that may be beneficial in the short term can affect the worker's trust in the system, which in turn influences whether future recommendations will be followed.
Motivated by this, in this paper, we formally model and analyze such interactions between worker's health, trust, and system decisions as a sequential decision problem.

In more detail, in Section~\ref{sec:model}, we propose a discrete-time Markov Decision Process (MDP) model of a trust-aware health recommender system interacting repeatedly with a worker whose health and trust evolve over time. Each worker has a two-dimensional type, reflecting their ``trust sensitivity'' and ``health sensitivity''. The recommender chooses policy parameters that affect when recommendations are triggered and how frequently they are issued to each worker, while the worker's trust affects the probability of following them. We then proceed to applying either a model-based approach (short-horizon, analytical) or a model-free approach (long-horizon, reinforcement learning) to finding the optimal, trust-aware health recommendation policy for each user type. 

Specifically, we first analytically characterize the recommender's two-step planning policy, and consider the role of the worker's starting health state, their type, and the recommender's focus on health vs. productivity, on this policy (Section~\ref{sec:analysis}). We then use the proposed framework to model and simulate recommender-worker interactions and to find an RL-based recommender, using Deep Q-Network (DQN), in the same setting (Section~\ref{sec:RL}). To do so, we develop and use a general PyTorch-based simulation framework for studying recommender-user interactions. We numerically compare the model-based and model-free policies (Section~\ref{sec:Discussion}). 

Through our analysis, we find that both policies can be used to build trust and improve worker health over time. That said, the advantage of long-term RL planning over short-horizon planning is most salient when dealing with workers with higher trust sensitivity. For such workers, a short-term policy avoids making recommendations for the fear of losing productivity without substantial health gains, whereas the RL policy identifies the potential for gradual trust building. We note that while our modeling is motivated by construction workplaces, these findings can have implications for trust-building in general health recommender systems.
\section{Related Work}\label{sec:related-work}

Monitoring workers' health and safety conditions in construction environments has been widely studied in prior work \cite{aryal2017monitoring, shakerian2021assessing, rasouli2024smart, ahn2019wearable, nnaji2020technologies}. However, relatively less attention has been given to how such information should be used to provide personalized feedback or adaptive interventions for workers. 

Some works have started to move in this direction. For example, \cite{shayesteh2024evaluating} studies the use of machine learning models and decision trees to provide personalized health-related feedback during routine construction activities. \cite{hu2023personalized} studies how safety interventions can be tailored across workers by incorporating cognitive-related factors into the personalization process. \cite{kim2020development} considers a system that tracks workers' physiological conditions in hot environments and includes a personal management component that alerts the manager or supervisor when the worker reaches a higher risk level. More broadly, beyond the construction workplace context, health recommender systems have been developed to provide personalized recommendations for improving health, and recent reviews summarize their development across a range of health applications \cite{de2021health, cai2022health}. All these works tackle personalized intervention design, but they do not account for the impact of trust, nor do they study the formal trust-aware sequential policy design considered in our paper. 

The importance of considering trust in recommendation systems has been identified empirically by works such as \cite{harman2014dynamics}, which examines the dynamics of user trust in recommender systems and shows that trust can change over repeated interaction depending on recommendation quality and user preferences. Trust in construction workplaces has mainly been studied in the context of human-AI and human-robot collaboration. \cite{chang2024toward} reviews this literature (trust-building in construction) and notes that trust in robots should not be viewed as fixed, but rather as something that can evolve over time and depend on the robot's performance. \cite{emaminejad2024assessing} studies trust in AI-powered collaborative robots (cobots) in construction and shows that issues such as safety and reliability can play an important role in whether workers are willing to use these systems. However, these works focus on robots, cobots, and collaborative AI systems in construction, rather than on trust in a health-oriented recommender system, and also lack the control-theoretic formalism of this paper.

Prior works \cite{chen2018planning, akash2020human, sprenger2024control, ravazzi2026optimal} have modeled evolving user response within sequential decision-making and control frameworks. These works have similarities to ours, since they also study how current system decisions affect future user response, but they do not study recommendations with the combination of \emph{endogenously-evolving} trust, trust-driven compliance, and the coupling between two different user states. In more detail, \cite{chen2018planning} and \cite{akash2020human} model trust as a dynamic quantity in human-machine interaction using partially observable Markov decision process (POMDP) formulation. In particular, \cite{akash2020human} relates trust to compliance, that is, whether the human follows the system's recommendation, which is close to our setting, but the difference remains that \cite{akash2020human} does not consider a direct coupling between trust and a second user state, such as health. In another related setting, \cite{sprenger2024control, ravazzi2026optimal} study how (content) recommendations should be designed in social networks by modeling user's evolving responses due to opinion dynamics. They consider both model-free and model-based control (MPC) strategies for maximizing user engagement, wherein user engagement is influenced by the user's opinion on \emph{what} is recommended; our setting is similar in spirit, in that compliance is impacted by trust in \emph{who} recommends; a key difference is that we also consider a second, coupled health state, the improvement of which (as opposed to mere engagement) is the ultimate goal of the recommendation system. 
\vspace{-0.08in}
\section{Model}\label{sec:model}

We begin by introducing a model of interaction between a health recommender system and a user (here, a construction worker). Interaction between the recommender and the user occurs at discrete decision times $t=1,2,\dots$, where each decision interval may represent, e.g., 1 hour.

\subsection{The state}

The state of the system at time $t$ is a two-dimensional vector $S_t=(H_t, T_t)$. Here, the first dimension $H_t \in [0, \bar{H}]$ denotes the user's health state at time $t$, with $\bar{H}$ representing the healthy baseline. The user's (physiological) data can be collected using wearable sensing devices, which can provide the health measurements needed to estimate the worker's state \cite{gautam2026memory}. The second dimension of the state,  $T_t \in [0,1]$, denotes the user's trust in the recommender at time $t$; this is the new information we are proposing should be tracked by the system. We also define the health loss at time $t$ as $\Delta H_t := \bar{H} - H_t$. The user's starting health and trust are $s_0=(\bar{H}, T_0)$.

\subsection{Recommender's and user's actions}

\paragraph{The recommender's action} At each time $t$, the recommender's policy determines whether to issue a recommendation or not. We focus on a single type of recommendation, such as ``seek shade for 10 minutes'' or ``take a break to drink water''; the specific recommendation type is fixed, and only its timing is controlled. The generated recommendation is sent to users through mobile applications or wearable devices \cite{shayesteh2024evaluating}. We denote this by $u_t \in \{0,1\}$, where $u_t = 1$ indicates that the recommendation is issued, and $u_t = 0$ otherwise.   

More specifically, at each decision time $t$, the recommender compares the current health loss $\Delta H_t$ with a decision threshold $\eta_t \in [0, \bar{H}]$. If $\Delta H_t \leq \eta_t$ (i.e., health degradation is assessed to be small), no recommendation is issued. If $\Delta H_t > \eta_t$ (i.e., significant drop in health is detected), then the recommender issues the recommendation with probability $f_t$, where $f_t\in [0,1]$ captures the recommendation frequency. Accordingly,
\[\mathbb{P}(u_t=1|\Delta H_t > \eta_t) = f_t \quad \mathbb{P}(u_t=1|\Delta H_t \leq \eta_t) = 0.\]
The recommender's action consists of choosing both the recommendation triggering threshold $\eta_t$ and the recommendation frequency $f_t$ at each time step.

\paragraph{The user's actions} After receiving a recommendation, the user may either follow or ignore it. We denote the user's \emph{compliance} by $c_t \in \{0,1\}$ where $c_t = 1$ denotes that the user follows the recommendation. We assume the user's response is shaped by their current trust level. Specifically, when a recommendation is issued, the user follows it with probability $T_t$; that is, $\mathbb{P}(c_t=1|u_t=1) = T_t$. Thus, a higher trust corresponds to a higher likelihood of compliance. If there is no recommendation, then $c_t =0$ by default. 
\vspace{-0.05in}
\subsection{State dynamics}
\paragraph{Health dynamics} The user's health evolves depending on whether the recommendation is followed. Specifically,
\vspace{-0.1in}
\begin{equation}\label{eq:health-update}
H_{t+1} = \max\{ 0, \bar{H} - \bigl(c_t \kappa_1 + (1-c_t)\kappa_2\bigr)\Delta H_t\},
\end{equation} 
where $\kappa_1$ and $\kappa_2$ characterize the evolution of the health loss under compliance and non-compliance, respectively, and are assumed to be user-independent. We will also assume $0 < \kappa_1 < 1 < \kappa_2$. Here, the choice of $\kappa_1<\kappa_2$ implies that recovery happens at a slower time-scale than it takes for the health degradation to accrue. This choice is motivated by findings from the construction health literature, where \cite{jaber2013incorporating} argues that in practice, taking rests does not usually lead to complete recovery. The choice of $\kappa_2>1$, on the other hand, implies that health degradation will progressively speed up if no recovery periods are taken.
\paragraph{Trust dynamics} The user's trust is updated based on the resulting health state after the interaction. In particular, the trust dynamics depend on whether the next step health loss crosses a \emph{health sensitivity} threshold $\delta_H \in [0, \bar{H}]$, and on the user's \emph{trust sensitivity} $\delta_T\in[0,1]$. Specifically,
\vspace{-0.03in}
\begin{equation}\label{eq:trust-update}
T_{t+1} = \min \{1, \max \{0,T_t - (-1)^{u_t} \mathrm{sign}(\Delta H_{t+1} - \delta_H) \delta_T \} \},
\end{equation}
where we adopt the convention $\mathrm{sign}(0):=1$. 
We let both $\delta_H$ and $\delta_T$ be user-specific parameters; together, they determine the \emph{user's type} $(\delta_H, \delta_T)$.

This update captures that the direction of trust change depends both on whether a recommendation is issued ($u_t=1$) and on whether the resulting health loss lies above or below the fixed health sensitivity threshold $\delta_H$. In particular, when a recommendation is issued, the trust update leads to increased trust in the system if the resulting health impact exceeds $\delta_H$, and decreases otherwise. Conversely, when no recommendation is issued, trust increases if the resulting health loss remains below the health sensitivity threshold, and decreases otherwise. 

Table~\ref{t:trust-dynamics} summarizes the different changes in trust. Note also that based on the earlier assumption of $\kappa_1 < \kappa_2$ on the health dynamics, the case in which the health loss remains below $\delta_H$ under non-compliance but exceeds it under compliance cannot occur simultaneously ($\dagger$ in Table~\ref{t:trust-dynamics}).
\vspace{-0.05in}
\begin{table}[h]
\centering
\caption{Trust update under different recommendation and compliance outcomes.}
\label{t:trust-dynamics}
\setlength{\tabcolsep}{4pt}
\begin{tabular}{c|c|c|c}
\hline
Condition & $(u_t,c_t)=(1,1)$ & $(u_t,c_t)=(1,0)$ & $(u_t,c_t)=(0,0)$ \\
\hline
$\Delta H_{t+1}\ge \delta_H$ 
& $T_t+\delta_T^{\dagger}$ 
& $T_t+\delta_T$ 
& $T_t-\delta_T$ \\
$\Delta H_{t+1}<\delta_H$ 
& $T_t-\delta_T$ 
& $T_t-\delta_T^{\dagger}$ 
& $T_t+\delta_T$ \\
\hline
\end{tabular}
\vspace{1mm}
{\footnotesize $^{\dagger}$ Cannot hold simultaneously under the assumption $\kappa_1 < \kappa_2 $.}
\vspace{-0.2in}
\end{table}

\subsection{Recommender's objective}

At each decision time $t$, the recommender observes the current user state $S_t=(H_t, T_t)$ (as well as potentially the history) and uses the policy parameters $(\eta_t, f_t)$ to decide whether to issue a recommendation. The parameters should be selected so as to maintain the user's health and productivity over time. Specifically, when the worker follows a recommendation, an instantaneous productivity cost $\delta_P > 0$ is incurred, reflecting time spent resting, hydrating, or otherwise stepping away from work. This leads to a per-time step loss of 
\begin{align*}
    l_t = \mathbb{E}[\lambda \Delta H_{t+1} + (1-\lambda) \delta_P c_t]
\end{align*}
where the expectation is taken over the randomness induced by the recommendation frequency as well as the user's stochastic compliance. Here, $\lambda \in [0,1]$ captures the trade-off between the importance of health loss and productivity loss from the recommender's perspective. 

Given this, the recommender's long-run optimization problem will be given by
\vspace{-0.15in}
\begin{align}
    \min ~~ &\mathbb{E}\big[\sum_{t=1}^\infty \gamma^{t-1} l_t\big]\notag\\
    ~~\text{s.t.} & ~~ \eta_t\in[0,\bar{H}], f_t\in [0,1]~.
    \label{eq:recommender-utility}
\end{align}
Here, $\gamma \in [0,1]$ is a standard discount factor. 

The minimization in \eqref{eq:recommender-utility} is over the space of all possible policies. In the following sections, we will proceed with solving this problem over two potential choices: (i) we first consider a short-horizon planning problem, where at each time $t$, the recommender chooses a policy $(\bar{\eta}_t, \bar{f}_t)$ based on the current state $S_t$ and while accounting for its impact over the next \emph{two} time steps. This will be used as a sliding window, and can be viewed as a simple Model-Predictive Control (MPC) approach; (ii) alternatively, we will use reinforcement learning (specifically, a DQN~\cite{mnih2015human}) to learn a policy that specifies $(\eta, f)$ while considering the long horizon problem. We will begin by describing the process for finding each policy (in Sections~\ref{sec:analysis} and~\ref{sec:RL}) followed by a comparison of the pros and cons of each approach and their implications for personalized health recommender systems (Section~\ref{sec:Discussion}). 
\section{Short-Horizon Analytical Policy Design}\label{sec:analysis}

We start by considering a recommender who truncates the optimization problem in \eqref{eq:recommender-utility} to a horizon of length two at a time. Formally, after observing the current state $S_t=(H_t, T_t)$, the recommender chooses a policy $(\bar{\eta}_t, \bar{f}_t)$ to be applied for the next two time steps, to minimize the cumulative loss: 
\begin{align*}
J(\bar{\eta}_t,\bar{f}_t)
&:=\mathbb{E}\Big[
\lambda\big(\Delta H_{t+1}+\gamma \Delta H_{t+2}\big) +(1-\lambda)\big(c_{t}+\gamma c_{t+1}\big)\delta_P
\Big].
\end{align*}
The dependence of $J$ on $\bar{\eta}_t$ and $\bar{f}_t$ is through the compliance variables $c_{t}$ and $c_{t+1}$, which in turn determine the subsequent health and trust states. 
This short-horizon approach will enable analytical insights and sensitivity analysis into how the recommendation triggering threshold and the recommendation frequency should be adjusted for different user types. 

We begin with analyzing the optimal threshold $\bar{\eta}_t$ for each user type $(\delta_H, \delta_T)$ and for a given frequency $\bar{f}_t$. For simplicity, we assume $\kappa_2^2 \Delta H_t \leq \bar{H}$ and $\delta_T < \min \{T_t, 1-T_t\}$ in the following analysis, so that the saturation terms in \eqref{eq:health-update} and \eqref{eq:trust-update} are not triggered.

\begin{proposition}\label{prop:optimal-eta}
Consider a user type $(\delta_H, \delta_T)$ with starting state $S_t=(H_t, T_t)$, and a fixed recommendation frequency $\bar{f}_t>0$. Assume $\Delta H_t>0$, $\lambda>0$, and $T_t>\gamma\bar f_t\Omega$, where $\Omega :=
\begin{cases}
T_t(T_t-\delta_T)+2\delta_T,
& \Delta H_t<\dfrac{\delta_H}{\kappa_2},\\[1mm]
T_t(T_t+\delta_T)-2\delta_T,
& \Delta H_t\geq\dfrac{\delta_H}{\kappa_2}.
\end{cases}$. Then, letting $\rho:=\frac{(1-\lambda)\delta_P}{\lambda(\kappa_2-\kappa_1)\Delta H_t}$, and $\kappa_{\text{mid}} := \frac{T_t(1+\gamma\kappa_2)-\gamma\kappa_2\bar{f}_t \Omega}{T_t - \gamma\bar{f}_t\Omega}$, if $\kappa_{\text{mid}} \in [\kappa_1, \kappa_2]$ the optimal recommendation triggering threshold $\bar{\eta}_t$ is
\begin{align*}
&\bar{\eta}_t \in
\begin{cases}
[0, \kappa_1 \Delta H_t) & \text{if } ~~ \rho\leq \kappa_1\\
[\kappa_1 \Delta H_t, \Delta H_t) & \text{if } ~~\kappa_1\leq \rho\leq \kappa_{\text{mid}}\\
[\Delta H_t, \kappa_2 \Delta H_t) & \text{if } ~~\kappa_{\text{mid}}\leq \rho\leq \kappa_2\\
[\kappa_2\,\Delta H_t, \bar{H}] & \text{if } ~~\kappa_2\leq \rho~.
\end{cases}
\end{align*} 
\end{proposition}

The proof sketch is provided in Appendix~\ref{app:proof-sketch}. When $\kappa_{\mathrm{mid}}\notin[\kappa_1,\kappa_2]$, one of the intermediate threshold regions is never optimal. Specifically, if $\kappa_{\mathrm{mid}}<\kappa_1$, the region $[\kappa_1\Delta H_t,\Delta H_t)$ is never optimal, whereas if
$\kappa_{\mathrm{mid}}>\kappa_2$, the region
$[\Delta H_t,\kappa_2\Delta H_t)$ is never optimal. The corresponding optimality conditions for these two cases and the complete proof of Proposition~\ref{prop:optimal-eta} are provided in the~\ref{app:online}.

\emph{Intuitive interpretation.} The optimal recommendation triggering threshold is guided by the parameter $\rho$, which looks at the relative importance of productivity loss vs. maintaining user health. Larger values of $\rho$ mean that the platform places more importance on productivity, or that the consequences of productivity loss are more costly. Accordingly, the platform can determine whether the optimal recommendation triggering threshold $\bar{\eta}_t$ should be selected to be low, medium-low, medium-high, or high. In particular, for larger values of $\rho$, the firm tends to pick higher recommendation triggering thresholds; i.e., when more focused on productivity, it is only when health degradation is severe that the platform considers issuing a recommendation. Finally, we note that the impact of a (given) recommendation frequency $\bar{f}_t$ on the optimal triggering threshold $\bar{\eta}_t$ depends on the users' trust dynamics as captured by $\Omega$: when $\Omega<0$ (baseline trust is fragile), lowering $\bar{f}_t$ pushes $\kappa_{\text{mid}}$ lower, making a higher triggering threshold optimal for the same $\rho$. Conversely, when $\Omega>0$, lowering $\bar{f}_t$ increases $\kappa_{\text{mid}}$ so that the platform compensates for the low frequency by triggering recommendations more easily.

In Section~\ref{sec:Discussion}, we evaluate the above expressions numerically, and use them to illustrate the selected recommender policies for each user type under this approach. Specifically, we consider a finite set of candidate values for $\bar{f}_t$, and identify the corresponding optimal $\bar{\eta}_t$ for each value from Proposition~\ref{prop:optimal-eta}. We then select the threshold-frequency pair that minimizes $J(\bar{\eta}_t, \bar{f}_t)$ as the final solution. In the numerical experiments, the saturation terms in~\eqref{eq:health-update} and~\eqref{eq:trust-update} are applied whenever needed.
\vspace{-0.08in}
\section{Long-Horizon RL-Based Policy Design}\label{sec:RL}

The previous analysis focused on the short-horizon setting, and by doing so, led to interpretable conditions for choosing the recommendation threshold. However, it failed to capture the long-run interaction between the recommender and the user, and their effects on health, trust, and productivity. We therefore also consider a model-free reinforcement learning (RL) formulation for the same MDP defined in Section~\ref{sec:model}, but which learns recommendation policies that optimize the cumulative discounted long-term reward in \eqref{eq:recommender-utility}.

To run our experiments, we developed a PyTorch-based simulator that tracks the interactions between the user and the recommender. Each simulation is initiated with the problem primitives, which include the user type ($\delta_H, \delta_T$), the user's health dynamic parameters ($\kappa_1,\kappa_2$), and the initial health and trust conditions; from these, only the initial health and trust are observable by the recommender. The simulation loop proceeds by drawing actions from the current trained policy, drawing realizations of the recommendation and user compliance, and tallying the associated reward accordingly; this loop is repeated for a fixed number of iterations. This simulator offers the capability to run experiments under different user types,  health sensitivities, horizon lengths, balance of health vs. productivity in the recommender cost, and varying initial health and trust conditions. It is available at \url{https://github.com/atefeh-mobr/personalized-trust-aware-health-recommender}.

\subsection{Simulation setup}
In the implementation used here, the training horizon is set to $T=30$. The environment parameters are initialized as $\bar H = 1, \kappa_1 = 0.7, \kappa_2 = 1.2, \lambda = 0.5, \delta_P = 0.2,$ 
with a discount factor of $\gamma=0.95$. 

We adopt a value-based reinforcement learning approach and approximate the action-value function using the standard DQN approach. To jointly optimize the continuous action pair $(\eta_t,f_t)$ with DQN, we discretize the admissible values. Specifically, the action space is defined as the Cartesian product of $\eta \in \{0.1,0.2,\dots,0.9\}$ and $f \in \{0, 0.25, 0.5, 0.75, 1\}.$
At the beginning of each training episode, initial conditions are sampled uniformly at random according to $H_0 \sim \mathrm{Unif}(0,\bar H), ~ T_0 \sim \mathrm{Unif}(0,1)$. 
The action-value function is approximated by a neural network with two hidden layers of width 128. Training is performed for 1000 episodes using Adam with learning rate $10^{-3}$, batch size 128, replay-buffer capacity $10^5$, soft target-network updates with $\tau=0.01$, and random seed 0. Exploration is implemented through an $\epsilon$-greedy rule, where $\epsilon$ is exponentially decayed from 0.4 to 0.05, and 30 replay-based parameter updates are performed after each episode.

After training, we evaluate the learned policy by generating simulated trajectories in the same simulator. For the cumulative reward comparison between the policies across different user types (Figure~\ref{fig:3_reward_diff}), the reported results are averaged over 20 trajectories for each user type to account for the stochasticity of the recommendation and compliance outcomes. For each trajectory, we record the cumulative reward together with the corresponding health, trust, recommendation, and compliance. For the trust and health evolution shown in Figure~\ref{fig:health_trust}, the reported trajectories are averaged over 1000 Monte Carlo simulations for each considered user type. Additional experiments examining the robustness to different initial conditions and RL training seeds are provided in the~\ref{app:online}.
\section{Discussion}\label{sec:Discussion}
We numerically illustrate the behavior of the short-horizon (two-step) and long-horizon (RL) policies. We focus on two aspects. First, we study how the selected recommendation triggering threshold  and recommendation frequency vary across user types under these two approaches. Second, we examine the dynamic behavior induced by these policies, and in particular, how the user's trust and health evolve.

\begin{figure*}[t]
    \centering
    \begin{subfigure}[t]{0.48\textwidth}
        \centering
        \includegraphics[width=1\textwidth]{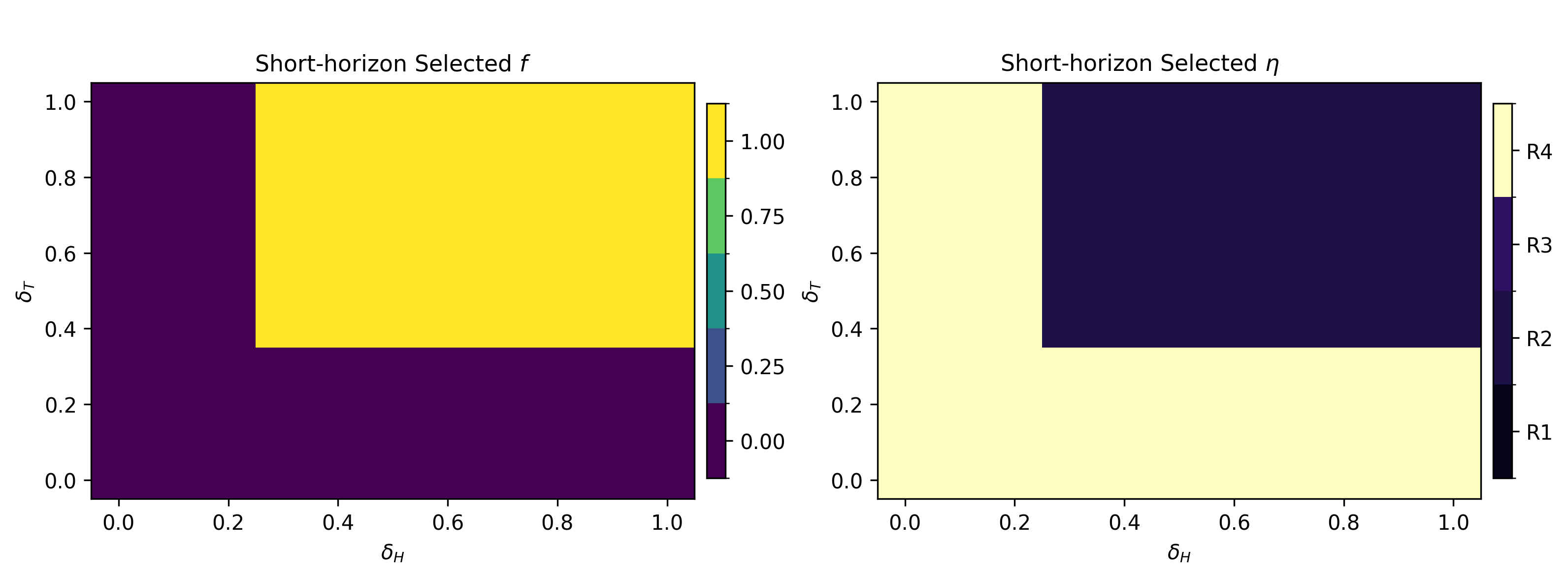}
        \caption{Selected recommendation frequency and triggering threshold $\eta$ under the short-horizon policy}
        \label{fig:1_analytic_policy_maps}
    \end{subfigure}
    \hfill
    \begin{subfigure}[t]{0.45\textwidth}
        \centering
        \includegraphics[width=1\textwidth]{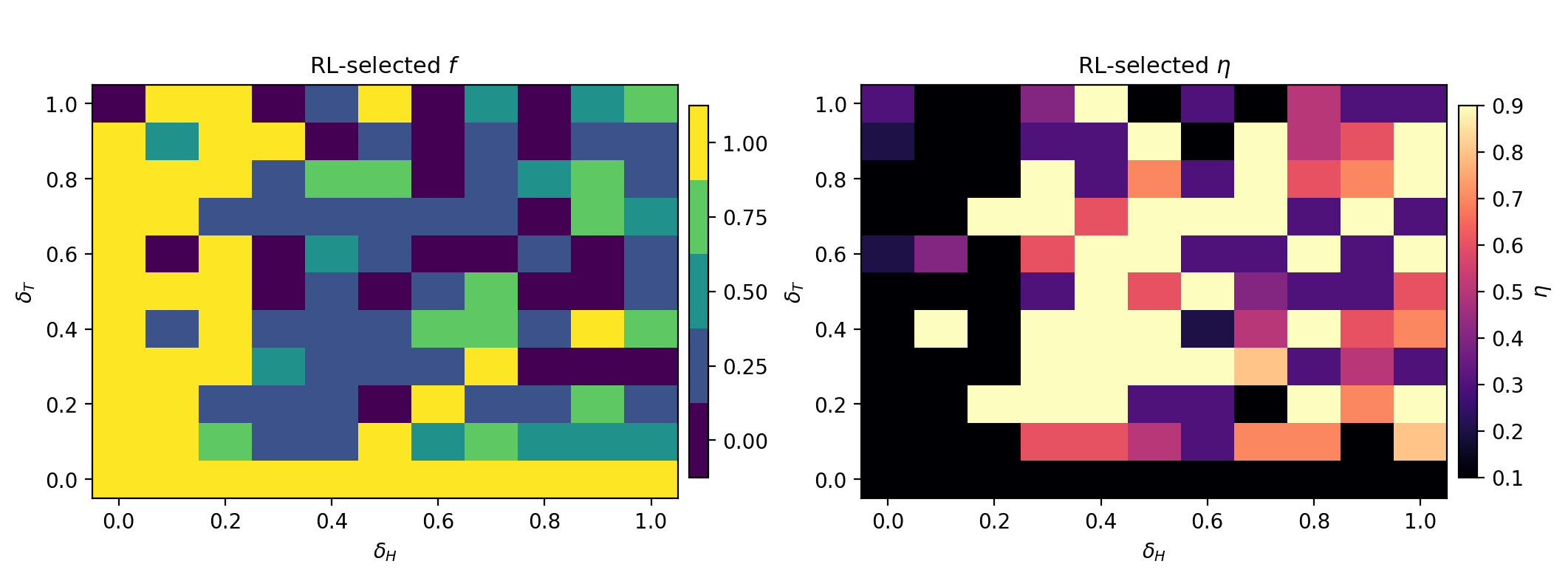}
        \caption{Selected recommendation frequency and triggering threshold $\eta$ under the RL policy}
        \label{fig:2_rl_policy_maps}
    \end{subfigure}
    \caption{Comparing short-horizon policy and RL in terms of frequency and decision threshold $\eta$ for different user types}
    \label{fig:compare-policies-sensitivity}
    \vspace{-0.2in}
\end{figure*}

\subsection{Comparison of the selected policies}
Figures~\ref{fig:1_analytic_policy_maps} and ~\ref{fig:2_rl_policy_maps} show the policy selected by the short-horizon model-based policy and the long-horizon RL policy, respectively, across different user types ($\delta_H, \delta_T$) over $[0,\bar{H}]\times[0,1]$. In all plots in this section, the current state is fixed at $(H_t, T_t)=(0.8, 0.5)$, leading to starting health degradation of $\Delta H_t=0.2$ (or 20\%). 

\paragraph{The short-horizon policy} Based on Figure~\ref{fig:1_analytic_policy_maps}, under this setting, the short-horizon policy separates the user space into two main regimes. For users with relatively small $\delta_T$, or with small $\delta_H$, the selected action is $(f=0,~ \eta\in[\kappa_2 \Delta H_t, \bar{H}])$, which effectively means that no recommendation would be generated for these user types. In contrast, for users with relatively large $\delta_T$ and $\delta_H$, the selected action is $(f=1, \eta\in[\kappa_1 \Delta H_t, \Delta H_t])$; that is, a recommendation is always triggered, and even for a relatively low initial health degradation.

This pattern can be understood from the way $\delta_H$ and $\delta_T$ affect the trust dynamics, and through trust, the expected health  and productivity loss from the recommender's view. Recall that $\delta_H$ determines whether the health change after a recommendation leads to a trust increase or decrease, while $\delta_T$ determines the magnitude of that trust update. 

When both $\delta_H$ and $\delta_T$ are small, recommending and not recommending lead to almost the same next step trust, meaning almost the same health loss. Therefore, the recommender prefers not to recommend, since the additional expected productivity cost from compliance does not overweigh the difference in health loss under compliance and non compliance. A similar behavior appears when $\delta_T$ is small even if $\delta_H$ is larger: since trust update remains small in magnitude, recommending does not create enough future health benefit to dominate its productivity cost.

When $\delta_H$ is small and $\delta_T$ is large, recommending can lead to a large trust increase, while not recommending decreases trust. Under this initialization, the higher trust increases expected compliance and thus the expected productivity cost. The short-horizon policy therefore, not accounting for the importance of maintaining trust, prefers not to recommend, as this cost dominates the corresponding health benefit.

Finally, when both $\delta_H$ and $\delta_T$ are large, recommending is beneficial, but the selected decision threshold is still chosen so that recommendations are not issued too aggressively, in particular after compliance at the first step. In other words, the chosen $\eta$-region allows a recommendation at the first step while avoiding unnecessary repeated recommendations over the two-step horizon.

\paragraph{The RL policy} Figure~\ref{fig:2_rl_policy_maps} shows the learned policies of the 121 discretely-sampled user-specific RL agents over the grid of user types. Compared to short-horizon policy, the overall trend is broadly reversed. In particular, for user types with relatively small $\delta_H$ or $\delta_T$, the RL agents generally select higher recommendation frequencies together with smaller values of $\eta$. In contrast, for user types with relatively large $\delta_H$ and $\delta_T$, they tend to choose lower frequencies together with larger threshold values. For medium $\delta_H$ and $\delta_T$ values, the learned policies again tend to use lower frequencies, but with even higher threshold values than in other regions.  Intuitively, over a long horizon, the RL agents learn that for users with either low $\delta_H$ or low $\delta_T$, recommending more frequently and at lower health-loss thresholds can be beneficial over time. On the other hand, for users in the upper-right region of the plots, the learned policies are generally more conservative: recommendations are issued less frequently and are typically triggered only when the health loss becomes sufficiently large. One possible explanation is that these users are more ``sensitive'', so more frequent unnecessary recommendations may reduce trust rapidly, hurting long horizon performance.

\subsection{Comparison of policy performance}

\begin{figure}[t]
    \centering
    \includegraphics[scale=0.32]{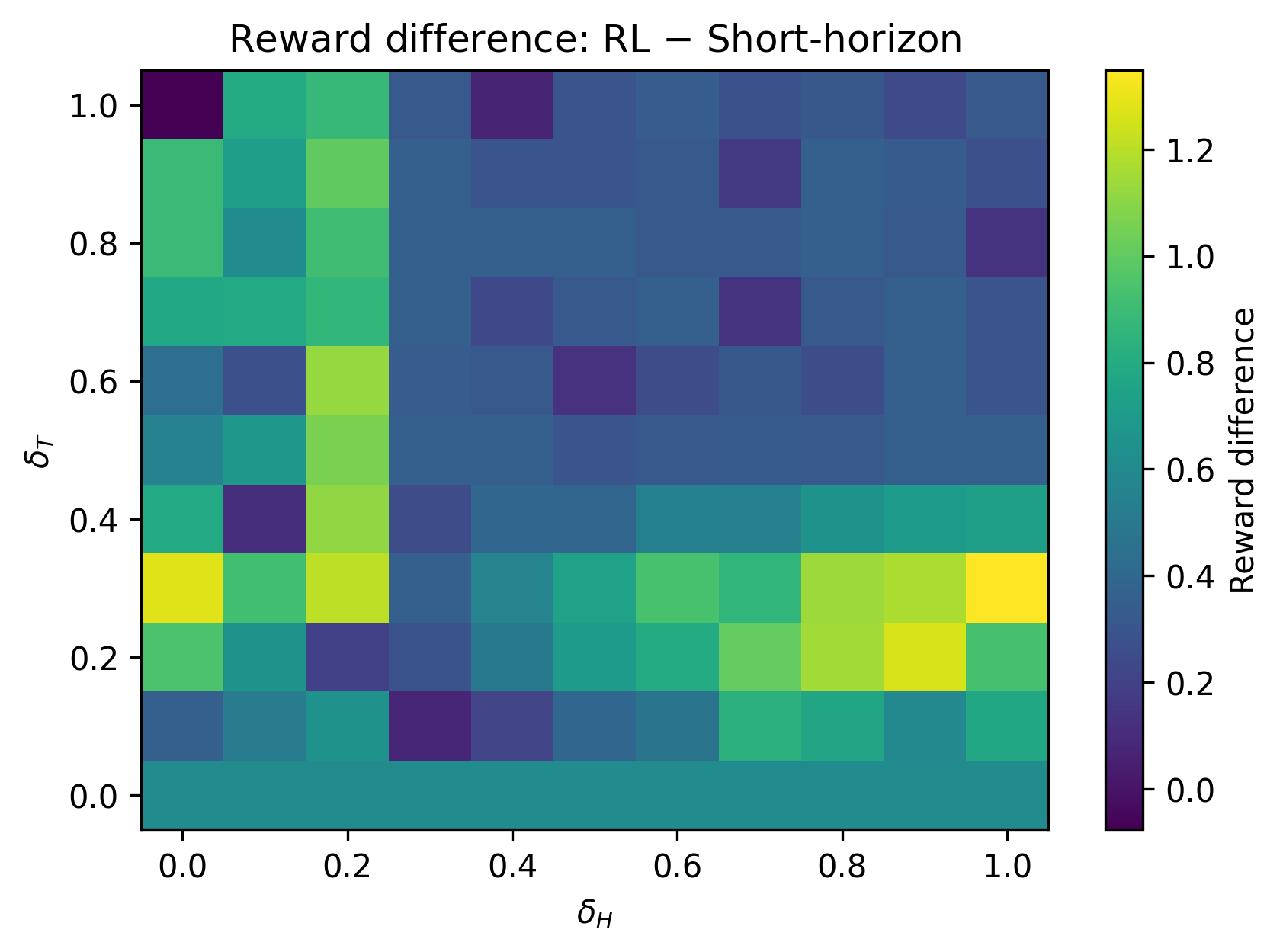}
    \caption{The difference in cumulative reward under short-horizon and RL policies for different user types}
    \label{fig:3_reward_diff}
    \vspace{-0.2in}
\end{figure}

As shown in Figure~\ref{fig:3_reward_diff}, the RL policy often outperforms the simpler short-horizon policy in terms of cumulative reward (defined as the negative of Eq.~\eqref{eq:recommender-utility}). That said, notably, despite its simplicity, the short-horizon policy remains competitive with, and at times outperforms, the RL-based planning. In particular, for user types with relatively large $\delta_H$ and $\delta_T$ (the upper-right side of the plot), the reward difference is small, which indicates that the two approaches are closer in performance. In contrast, for other user types, the reward difference is more noticeable. This is particularly visible in regions where the short-horizon policy tends not to recommend, while the RL policies opt for more aggressive recommendations with higher frequencies. This suggests that, over a long horizon, recommending more actively can provide additional benefit for users who are less sensitive to recommendation and trust updates, an effect not fully captured in short-horizon planning.

\subsection{Evolution of trust and health}
\begin{figure*}[t]
    \centering
    \begin{subfigure}[t]{0.48\textwidth}
        \centering
        \includegraphics[width=1\textwidth]{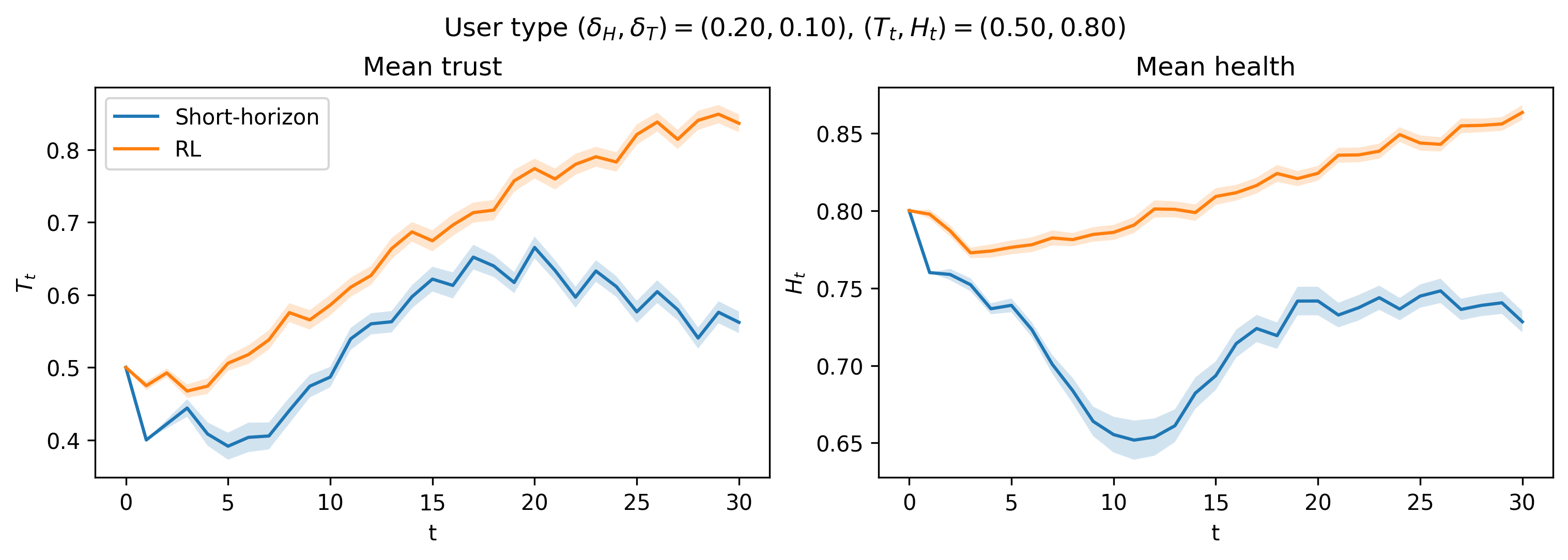}
        \caption{User type  $(\delta_H, \delta_T) = (0.2, 0.1)$}
        \label{fig:4_health-trust_low}
    \end{subfigure}
    \hfill
    \begin{subfigure}[t]{0.48\textwidth}
        \centering
        \includegraphics[width=1\textwidth]{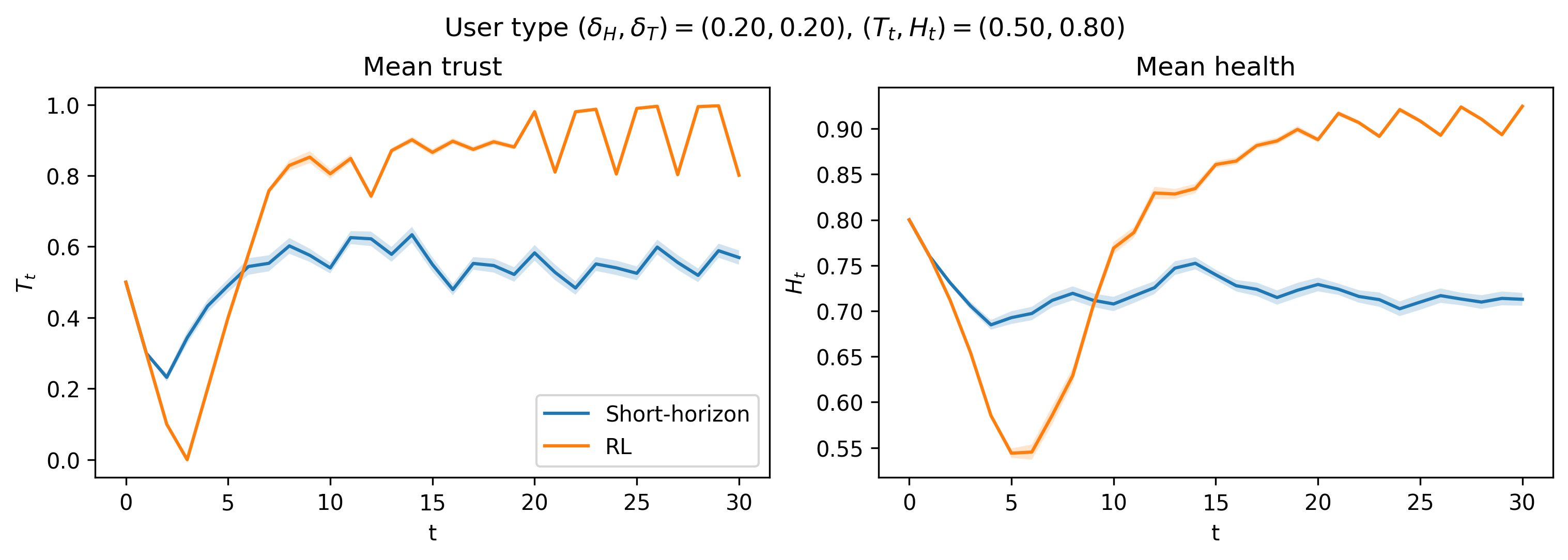}
        \caption{User type  $(\delta_H, \delta_T) = (0.2, 0.2)$}
        \label{fig:5_health-trust_high}
    \end{subfigure}
    \caption{Evolution of trust and health under different policies for different user types}
    \label{fig:health_trust}
    \vspace{-0.3in}
\end{figure*}

Figure~\ref{fig:health_trust} shows how trust and health evolve under both short-horizon and RL policies for two user types, $(\delta_H, \delta_T) = (0.2, 0.1)$ and $(\delta_H, \delta_T) = (0.2, 0.2)$, respectively.

In Figure~\ref{fig:4_health-trust_low}, we see that the short-horizon policy decides not to recommend early on as this user's type has low $\delta_H$ and $\delta_T$. At some point in time, by not recommending at all, it loses the user's trust, which leads to a drop in the user's health, which eventually leads the short-horizon policy to recommend, once health degradation is too high, in order to improve health. From that point on, both health and trust improve. On the other hand, the RL policy learns that recommending from the beginning can gradually build trust, as each trust update is small, and therefore improves health consistently. 

Figure~\ref{fig:5_health-trust_high} shows that for a user type with the same health sensitivity but higher trust sensitivity $\delta_T$, the policies behave differently. The short-horizon policy still sticks to not recommending, but the trust drop is sharper than before, so it changes its decision and begins recommending again. In contrast, RL again starts with recommending, which, in the early time steps, leads to lower trust and health compared to the short-horizon policy. This is only suboptimal in the short-term, though, as RL is planning to build trust and improve health in the long horizon, and eventually outperforms the short-horizon policy.
\section{Conclusion}\label{sec:conclusions}
This paper proposed a trust-aware dynamic model for personalized health recommendations in construction workplaces. The model captures the interaction between worker health, trust, compliance, and recommendation decisions over time. Our analysis provides insight into how recommendation thresholds and frequencies can be selected to better balance health outcomes and productivity, and suggests broader implications for trust-aware health recommender systems. In particular, we find that trust building can happen under both short-horizon and long-horizon recommendation policies, but it can be leveraged more effectively under long-horizon policies. Future work includes extending our model (e.g., to a POMDP where health and trust states are not fully observable) and also testing the simulator with construction workers in a lab environment (as in our earlier work on an agentic-AI recommendation framework~\cite{gautam2026memory}) to assess its effectiveness.

\bibliographystyle{IEEEtran}
\bibliography{references}

\appendix
\subsection{Proof Sketch of Proposition~\ref{prop:optimal-eta}}\label{app:proof-sketch}

To obtain the optimal threshold, we begin by determining the attainable two-step loss for a given user type and recommendation frequency, as shown in the Lemma~\ref{lemma:two-step-utility-by-region} below. 

\begin{lemma}\label{lemma:two-step-utility-by-region}
Consider a user type $(\delta_H, \delta_T)$ and a fixed recommendation frequency $\bar{f}_t$. Then, given the current state $S_t=(H_t, T_t)$, the optimal recommendation triggering threshold $\bar{\eta}_t$ is determined by $\underset{\eta}{\arg\min} ~\{J_1, J_2, J_3, J_4\}$
where
\[
\text{if } \quad \eta \ge \kappa_2 \Delta H_t
\quad\Longrightarrow\quad J_4=\lambda (1+\gamma\kappa_2)\kappa_2\,\Delta H_t,
\]
\begin{align*}
&\text{if } \quad \Delta H_t\le \eta<\kappa_2 \Delta H_t
\quad\Longrightarrow\\
&J_3=\begin{cases}
\lambda \Phi_{+}\kappa_2\Delta H_t + (1-\lambda)\gamma \bar{f}_t (T_t+\delta_T) \delta_P,
& \text{if } \Delta H_t<\dfrac{\delta_H}{\kappa_2},
\\[2mm]
\lambda\Phi_{-}\kappa_2\Delta H_t + (1-\lambda)\gamma \bar{f}_t (T_t-\delta_T) \delta_P,
& \text{if } \Delta H_t\ge\dfrac{\delta_H}{\kappa_2},
\end{cases}
\end{align*}

\begin{align*}
&\text{if } \quad \kappa_1 \Delta H_t \le \eta<\Delta H_t
\quad\Longrightarrow\quad\\
&J_2=\begin{cases}
\begin{aligned}
&\lambda \Big((1-\bar{f}_t)\Phi_{+}\kappa_2
+\bar{f}_t(1-T_t)\Phi_{-}\kappa_2\\
&\quad +\bar{f}_tT_t(1+\gamma\kappa_2)\kappa_1\Big)\,\Delta H_t\\
& + (1-\lambda) \bar{f}_t \Big(T_t + \gamma \big((1-\bar{f}_t) (T_t+\delta_T)\\
&\quad + \bar{f}_t (1-T_t)(T_t - \delta_T) \big)\Big)\delta_P,
\end{aligned}
& \text{if } \Delta H_t<\dfrac{\delta_H}{\kappa_2},
\\[15mm]
\begin{aligned}
&\lambda\Big((1-\bar{f}_t)\Phi_{-}\kappa_2
+\bar{f}_t(1-T_t)\Phi_{+}\kappa_2\\
&\quad +\bar{f}_tT_t(1+\gamma\kappa_2)\kappa_1\Big)\,\Delta H_t\\
& + (1-\lambda)\bar{f}_t \Big(T_t + \gamma \big((1-\bar{f}_t) (T_t-\delta_T)\\
&\quad + \bar{f}_t (1-T_t)(T_t + \delta_T) \big)\Big)\delta_P,
\end{aligned}
& \text{if } \Delta H_t\ge\dfrac{\delta_H}{\kappa_2},
\end{cases}
\end{align*}

\begin{align*}
&\text{if } \quad \eta<\kappa_1 \Delta H_t
\quad\Longrightarrow\quad\\
&J_1=
\begin{cases}
\begin{aligned}
&\lambda \Big((1-\bar{f}_t)\Phi_{+}\kappa_2
+\bar{f}_t(1-T_t)\Phi_{-}\kappa_2\\
&\quad +\bar{f}_tT_t\Phi_{-}\kappa_1\Big)\,\Delta H_t\\
& + (1-\lambda) \bar{f}_t\Big( T_t + \gamma \big((1-\bar{f}_t) (T_t+\delta_T)\\
&\quad + \bar{f}_t (T_t - \delta_T) \big)\Big)\delta_P,
\end{aligned}
\text{if } \Delta H_t<\dfrac{\delta_H}{\kappa_2},
\\[15mm]
\begin{aligned}
&\lambda \Big((1-\bar{f}_t)\Phi_{-}\kappa_2
+\bar{f}_t(1-T_t)\Phi_{+}\kappa_2\\
&\quad +\bar{f}_tT_t\Phi_{-}\kappa_1\Big)\,\Delta H_t\\
& + (1-\lambda) \bar{f}_t\Big(T_t + \gamma
\big((T_t-\delta_T)\\
&\quad + 2 \bar{f}_t\delta_T(1-T_t)\big)\Big)\delta_P,
\end{aligned}
\text{if } \dfrac{\delta_H}{\kappa_2}\le \Delta H_t<\dfrac{\delta_H}{\kappa_1},
\\[15mm]
\begin{aligned}
&\lambda\Big((1-\bar{f}_t)\Phi_{-}\kappa_2
+\bar{f}_t(1-T_t)\Phi_{+}\kappa_2\\
&\quad +\bar{f}_tT_t\Phi_{+}\kappa_1\Big)\,\Delta H_t\\
& + (1-\lambda) \bar{f}_t\Big(T_t + \gamma \big((1-\bar{f}_t) (T_t-\delta_T)\\
&\quad + \bar{f}_t(T_t + \delta_T)\big)\Big)\delta_P,\\
\end{aligned}
\text{if } \Delta H_t\ge\dfrac{\delta_H}{\kappa_1},
\end{cases}
\end{align*}

and \(
\Phi_{+}:=1+\gamma\Big(\kappa_2+\bar{f}_t(T_t+\delta_T)(\kappa_1-\kappa_2)\Big)\),  \(\Phi_{-}:=1+\gamma\Big(\kappa_2+\bar{f}_t(T_t-\delta_T)(\kappa_1-\kappa_2)\Big)\).

\end{lemma}

The proof of this lemma proceeds by expanding the two-step evolution of health and trust from the starting state according to the trust and health dynamic models and the stochasticity of the recommendation and the user's compliance. From there, using the expressions in Lemma~\ref{lemma:two-step-utility-by-region}, we proceed to comparing the four sub-costs $J_k$ under different conditions on $\Delta H_t$, and arrive at the conditions in Proposition~\ref{prop:optimal-eta} under which each of them is lowest.

\clearpage
\onecolumn
\section*{Online Appendix}
\refstepcounter{onlineappendix}
\label{app:online}

\setcounter{subsection}{0}
\renewcommand{\thesubsection}{\Alph{subsection}}
\subsection{Proof and Extensions of Proposition~\ref{prop:optimal-eta}}

Now, using Lemma~\ref{lemma:two-step-utility-by-region}, for fixed
$\bar{f}_t$, it suffices to compare $J_1,J_2,J_3,$ and $J_4$.

Using the definition of $\rho$, direct simplification gives
\[J_2-J_1=\lambda\gamma\bar{f}_t^2(\kappa_2-\kappa_1)\Delta H_t
(\kappa_1-\rho)
\begin{cases}
T_t(T_t-\delta_T),
& \Delta H_t<\dfrac{\delta_H}{\kappa_1},\\[1mm]
T_t(T_t+\delta_T),
& \Delta H_t\geq\dfrac{\delta_H}{\kappa_1}.
\end{cases}\]
Hence, $J_1\leq J_2 \iff \rho\leq\kappa_1$. 

Similarly, using the definitions of $\Omega$ and
$\kappa_{\text{mid}}$,
\[J_3-J_2 = \lambda\bar{f}_t(\kappa_2-\kappa_1)\Delta H_t
(T_t-\gamma\bar{f}_t\Omega)
(\kappa_{\text{mid}}-\rho).\]
Given $T_t>\gamma\bar{f}_t\Omega$, we have $J_2\leq J_3 \iff \rho\leq\kappa_{\text{mid}}$. 

Finally,
\[J_4-J_3 =
\lambda\gamma\bar{f}_t(\kappa_2-\kappa_1)\Delta H_t
(\kappa_2-\rho)
\begin{cases}
T_t+\delta_T,
& \Delta H_t<\dfrac{\delta_H}{\kappa_2},\\[1mm]
T_t-\delta_T,
& \Delta H_t\geq\dfrac{\delta_H}{\kappa_2},
\end{cases}\]
and therefore $J_3\leq J_4 \iff \rho\leq\kappa_2$. 

Below, we also extend Proposition~\ref{prop:optimal-eta} when $\kappa_{\text{mid}} \notin [\kappa_1, \kappa_2]$. 

\begin{corollary}\label{cor:optimal-eta-lower}
Under the settings of Proposition~\ref{prop:optimal-eta}, if $\kappa_{\text{mid}} < \kappa_1$, the threshold region $[\kappa_1 \Delta H_t, \Delta H_t)$ is never optimal, and the optimal recommendation triggering threshold $\bar{\eta}_t$ satisfies
\begin{align*}
&\bar{\eta}_t \in
\begin{cases}
[0, \kappa_1 \Delta H_t) & \text{if } ~~ \rho\leq \kappa_{13}\\
[\Delta H_t, \kappa_2 \Delta H_t) & \text{if } ~~\kappa_{13}\leq \rho\leq \kappa_2\\
[\kappa_2\,\Delta H_t, \bar{H}] & \text{if } ~~\kappa_2\leq \rho\\
\end{cases}
\end{align*}

where
\begin{itemize}

\item \textbf{Case $1$.} When $\Delta H_t<\frac{\delta_H}{\kappa_2}$, then $\kappa_{13} = \frac{T_t(1+\gamma\kappa_2)-\gamma\kappa_2\bar{f}_t \Omega_1 + \gamma \bar{f}_t T_t (T_t - \delta_T)\kappa_1}{T_t - \gamma\bar{f}_t\Omega_1+ \gamma \bar{f}_t T_t (T_t - \delta_T)}$.

\item \textbf{Case $2$.} When $\frac{\delta_H}{\kappa_2}\leq \Delta H_t<\frac{\delta_H}{\kappa_1}$, then $\kappa_{13} = \frac{T_t(1+\gamma\kappa_2)-\gamma\kappa_2\bar{f}_t \Omega_2 + \gamma \bar{f}_t T_t (T_t - \delta_T)\kappa_1}{T_t - \gamma\bar{f}_t\Omega_2 + \gamma \bar{f}_t T_t (T_t - \delta_T)}$.

\item \textbf{Case $3$.} When $\frac{\delta_H}{\kappa_1}\leq \Delta H_t$, then $\kappa_{13} = \frac{T_t(1+\gamma\kappa_2)-\gamma\kappa_2\bar{f}_t \Omega_3 + \gamma \bar{f}_t T_t (T_t + \delta_T)\kappa_1}{T_t - \gamma\bar{f}_t\Omega_3 + \gamma \bar{f}_t T_t (T_t + \delta_T)}$

\end{itemize}
\end{corollary} 

\begin{corollary}\label{cor:optimal-eta-upper}
Under the settings of Proposition~\ref{prop:optimal-eta}, if $\kappa_{\text{mid}} > \kappa_2$, the threshold region $[\Delta H_t, \kappa_2 \Delta H_t)$ is never optimal, and the optimal recommendation triggering threshold $\bar{\eta}_t$ satisfies
\begin{align*}
&\bar{\eta}_t \in
\begin{cases}
[0, \kappa_1 \Delta H_t) & \text{if } ~~ \rho\leq \kappa_{1}\\
[\kappa_1 \Delta H_t, \Delta H_t) & \text{if } ~~\kappa_{1}\leq \rho\leq \kappa_{24}\\
[\kappa_2\,\Delta H_t, \bar{H}] & \text{if } ~~\kappa_{24}\leq \rho\\
\end{cases}
\end{align*}

where
\begin{itemize}

\item \textbf{Case $1$.} When $\Delta H_t<\frac{\delta_H}{\kappa_2}$, then $\kappa_{24} = \frac{T_t(1+\gamma\kappa_2)-\gamma\kappa_2\bar{f}_t \Omega_1 + \gamma (T_t + \delta_T)\kappa_2}{T_t - \gamma\bar{f}_t\Omega_1+ \gamma (T_t + \delta_T)}$.

\item \textbf{Case $2$.} When $\frac{\delta_H}{\kappa_2}\leq \Delta H_t<\frac{\delta_H}{\kappa_1}$, then $\kappa_{24} = \frac{T_t(1+\gamma\kappa_2)-\gamma\kappa_2\bar{f}_t \Omega_2 + \gamma (T_t - \delta_T)\kappa_2}{T_t - \gamma\bar{f}_t\Omega_2 + \gamma  (T_t - \delta_T)}$.

\item \textbf{Case $3$.} When $\frac{\delta_H}{\kappa_1}\leq \Delta H_t$, then $\kappa_{24} = \frac{T_t(1+\gamma\kappa_2)-\gamma\kappa_2\bar{f}_t \Omega_3 + \gamma (T_t - \delta_T)\kappa_2}{T_t - \gamma\bar{f}_t\Omega_3 + \gamma (T_t - \delta_T)}$

\end{itemize}
\end{corollary}

\subsection{Additional Numerical Illustration of the Short-Horizon Policy}

\begin{figure*}[h]
    \centering
    \begin{subfigure}[t]{0.48\textwidth}
        \centering
        \includegraphics[width=1\textwidth]{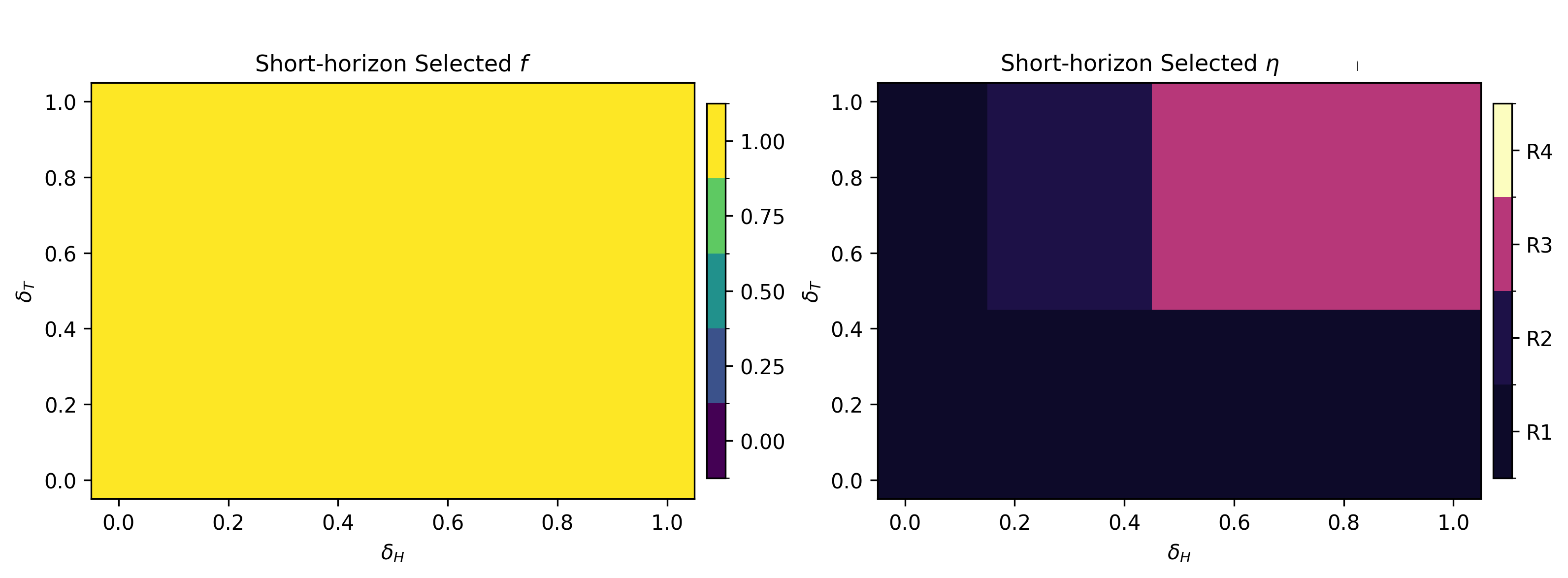}
        \caption{Short-horizon policy}
        \label{fig:6_analytic_policy_maps_prop1}
    \end{subfigure}
    \hfill
    \begin{subfigure}[t]{0.48\textwidth}
        \centering
        \includegraphics[width=1\textwidth]{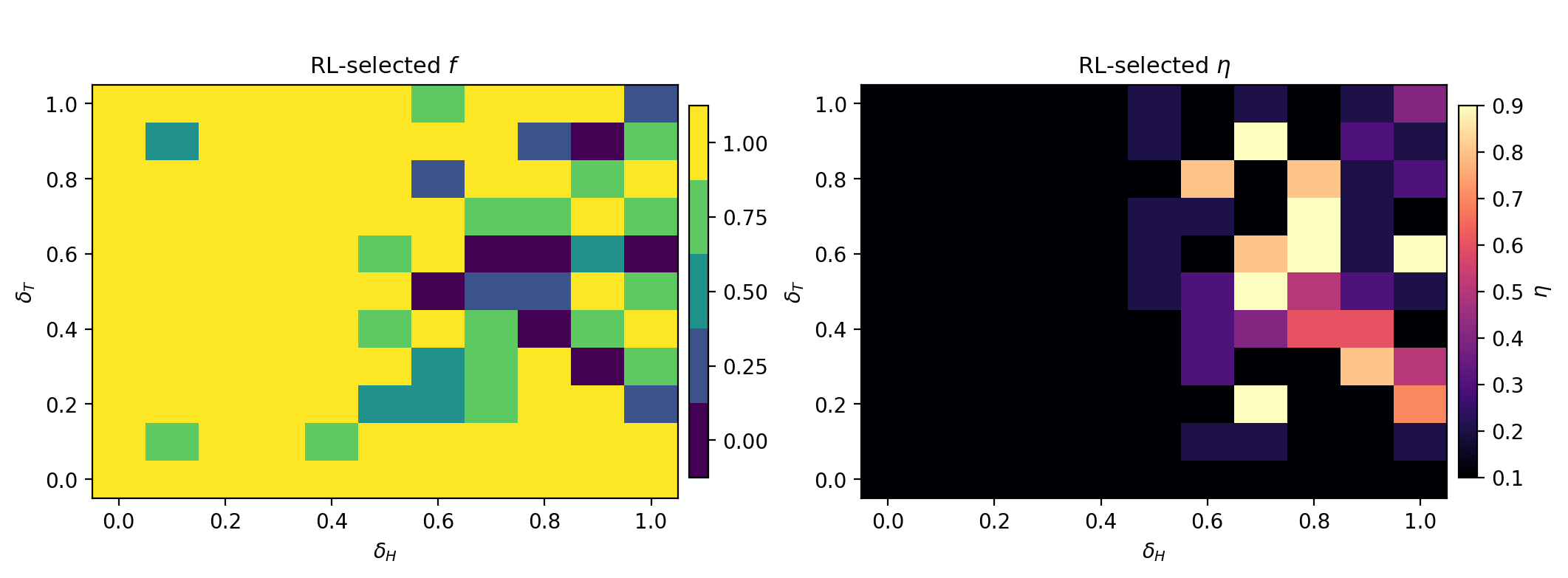}
        \caption{RL policy}
    \label{fig:7_rl_policy_maps_prop1}
    \end{subfigure}
    \caption{Comparing short-horizon policy and RL in terms of frequency and decision threshold $\eta$ for different user types, for a parameter setting illustrating the main threshold-region case considered in Proposition~\ref{prop:optimal-eta}.}
    \label{fig:policy-prop1}
    \vspace{-0.2in}
\end{figure*}

\begin{figure*}[t]
    \centering
    \begin{subfigure}[t]{0.48\textwidth}
        \centering
        \includegraphics[width=1\textwidth]{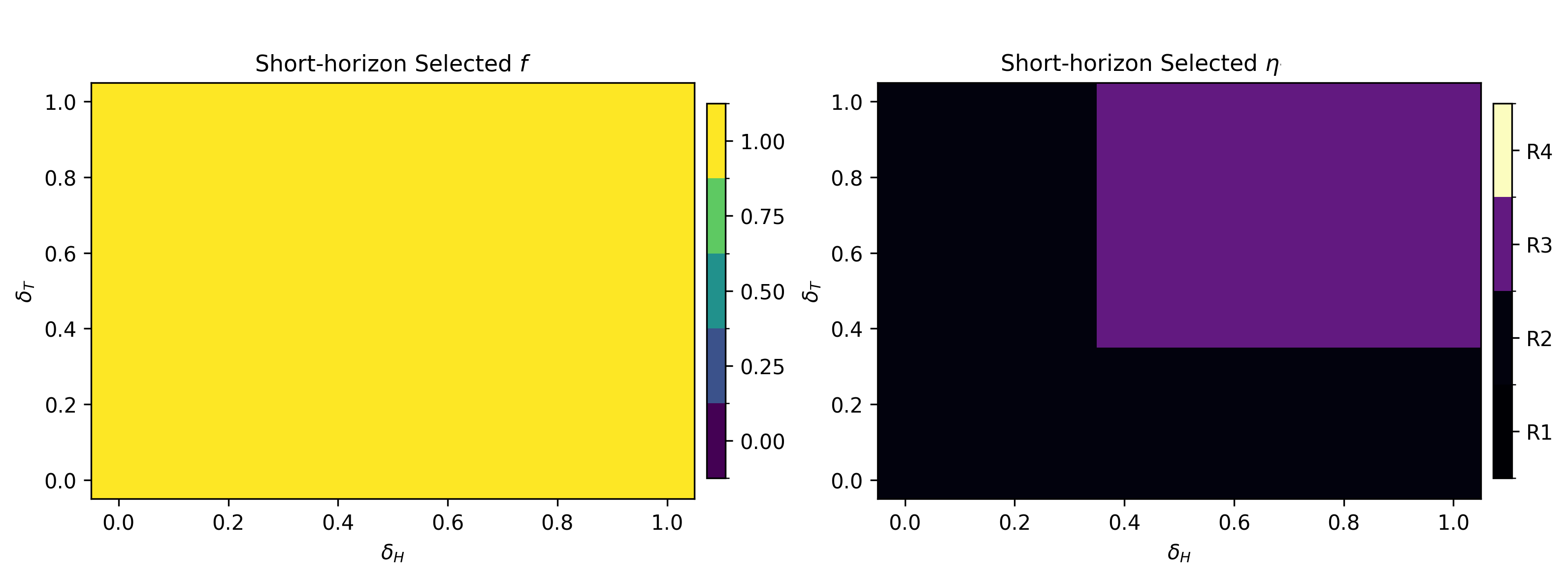}
        \caption{Short-horizon policy}
        \label{fig:8_analytic_policy_maps_cor1}
    \end{subfigure}
    \hfill
    \begin{subfigure}[t]{0.48\textwidth}
        \centering
        \includegraphics[width=1\textwidth]{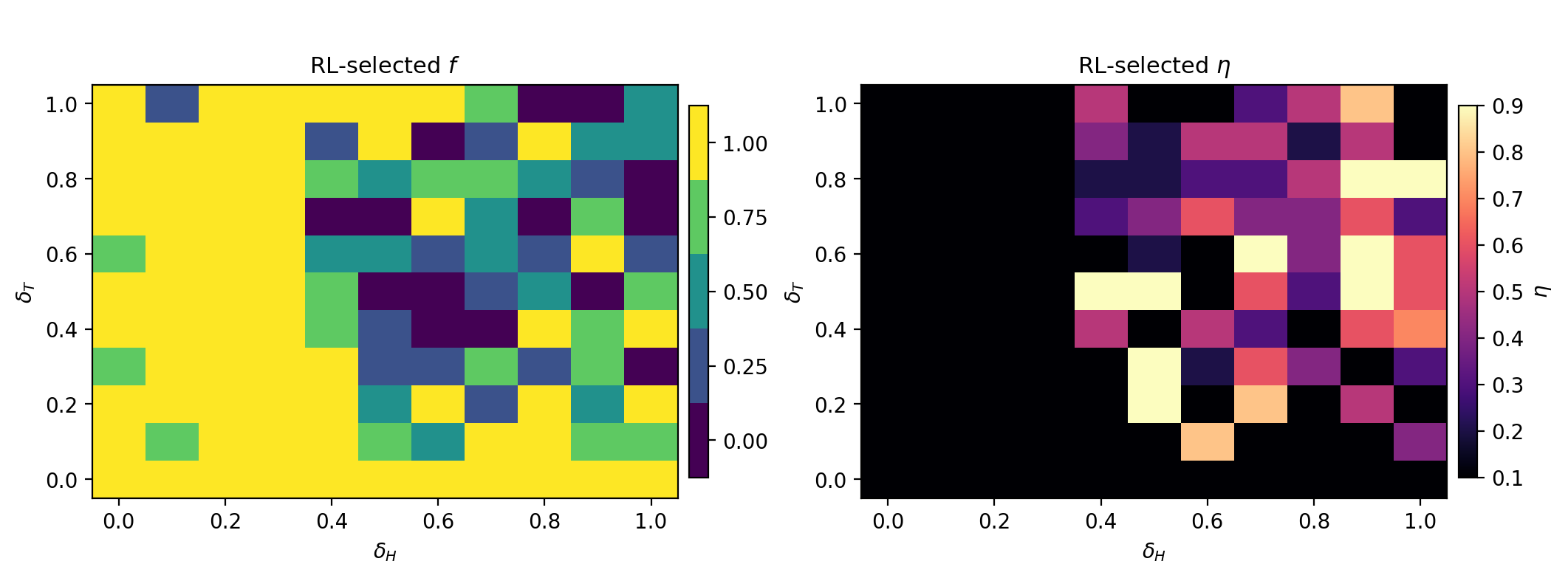}
        \caption{RL policy}
    \label{fig:9_rl_policy_maps_cor1}
    \end{subfigure}
    \caption{Comparing short-horizon policy and RL in terms of frequency and decision threshold $\eta$ for different user types, for a parameter setting
    corresponding to Corollary~\ref{cor:optimal-eta-lower} and Proposition~\ref{prop:optimal-eta}}
    \label{fig:policy-cor1}
\end{figure*}

Figures~\ref{fig:policy-prop1} and~\ref{fig:policy-cor1} illustrate the short-horizon policy and RL under two different parameter settings. In Figure~\ref{fig:policy-prop1}, the setting $(T_t,\Delta H_t)=(0.5,0.2), \kappa_1=0.8, \kappa_2 = 2.5, \gamma = 0.5$, corresponds to the main threshold-region case considered in Proposition~\ref{prop:optimal-eta}, with saturation effects present. Figure~\ref{fig:6_analytic_policy_maps_prop1} illustrates how the short-horizon policy changes the selected threshold region as the user type varies.
In Figure~\ref{fig:policy-cor1}, the setting $(T_t,\Delta H_t)=(0.5,0.12), \kappa_1=0.8, \kappa_2 = 2.8, \gamma = 0.5$, satisfies the conditions of Corollary~\ref{cor:optimal-eta-lower} for some user types, while the remaining user types satisfy the conditions of Proposition~\ref{prop:optimal-eta}. Figure~\ref{fig:8_analytic_policy_maps_cor1} again shows the changes in the selected threshold region across user types, while the region $[\kappa_1 \Delta H_t, \Delta H_t)$ does not appear for the user types corresponding to Corollary~\ref{cor:optimal-eta-lower}. In both examples, the short-horizon policy selects the highest recommendation frequency, $f=1$.

On the other hand, the RL policies in Figures~\ref{fig:7_rl_policy_maps_prop1} and~\ref{fig:9_rl_policy_maps_cor1} show more broadly similar qualitative patterns across the two parameter settings. For smaller $\delta_H$, RL generally selects higher recommendation frequencies, often $f=1$, and lower triggering thresholds, while for larger $\delta_H$ its decisions become more conservative and heterogeneous across user types. Therefore, for some user types with small $\delta_H$, the RL and short-horizon policies are more closely aligned, whereas outside these regions, the RL policy adjusts the recommendation frequency across user types while the short-horizon policy selects a single frequency, $f=1$, throughout. This is consistent with the long-horizon behavior observed in the main experiments, where recommending more frequently can be beneficial for less sensitive users, while more cautious recommendations can help avoid unfavorable trust dynamics for more sensitive users.

\subsection{Sensitivity to Initial Health and Trust}

Throughout this section, to study whether the policy patterns discussed in Section~\ref{sec:Discussion} are specific to the initial state used there, we repeat the comparison under three alternative initial health and trust states. All other parameters are unchanged, and the same trained RL agents (seed 0) are used throughout. The three initial states are: (i) high health and low trust $(H_t,T_t)=(0.9,0.2)$, (ii) low health and low trust $(H_t,T_t)=(0.5,0.2)$, (iii) high health and higher trust $(H_t,T_t)=(0.95,0.5)$.

\begin{figure*}[h]
    \centering
    \begin{subfigure}[t]{0.48\textwidth}
        \centering
        \includegraphics[width=1\textwidth]{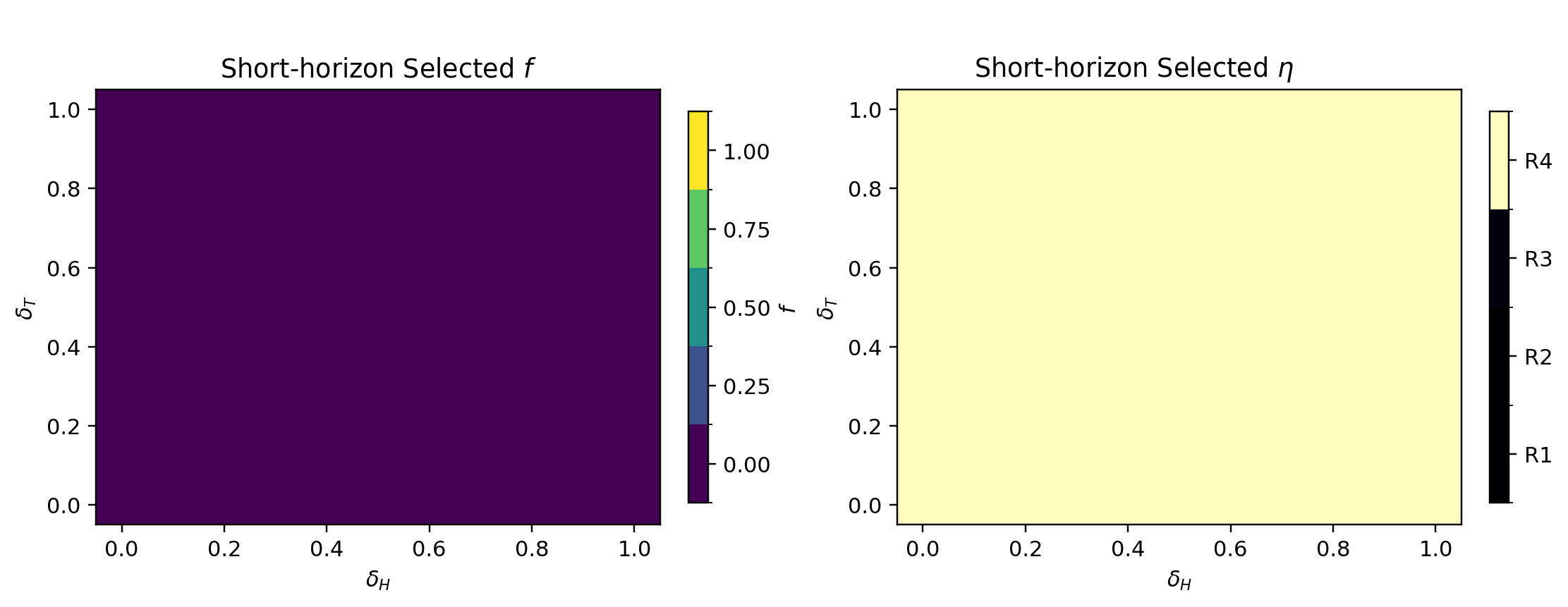}
        \caption{Short-horizon: $(H_t,T_t)=(0.9,0.2)$.}
        \label{fig:10_analytic_policy_low_deficit_low_trust_H0.9_T0.2}
    \end{subfigure}
    \hfill
    \begin{subfigure}[t]{0.48\textwidth}
        \centering
        \includegraphics[width=1\textwidth]{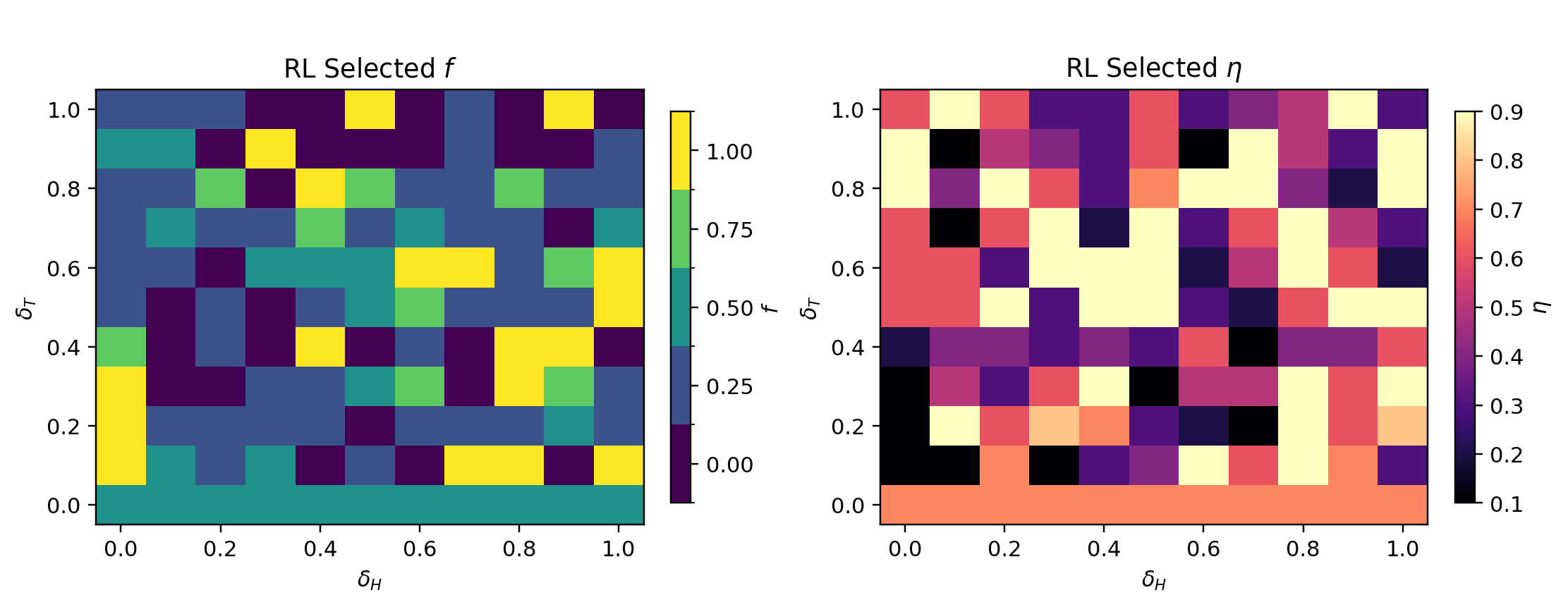}
        \caption{RL: $(H_t,T_t)=(0.9,0.2)$. }
    \label{fig:11_rl_policy_seed0_low_deficit_low_trust_H0.9_T0.2}
    \end{subfigure}
    \hfill
    \begin{subfigure}[t]{0.48\textwidth}
        \centering
        \includegraphics[width=1\textwidth]{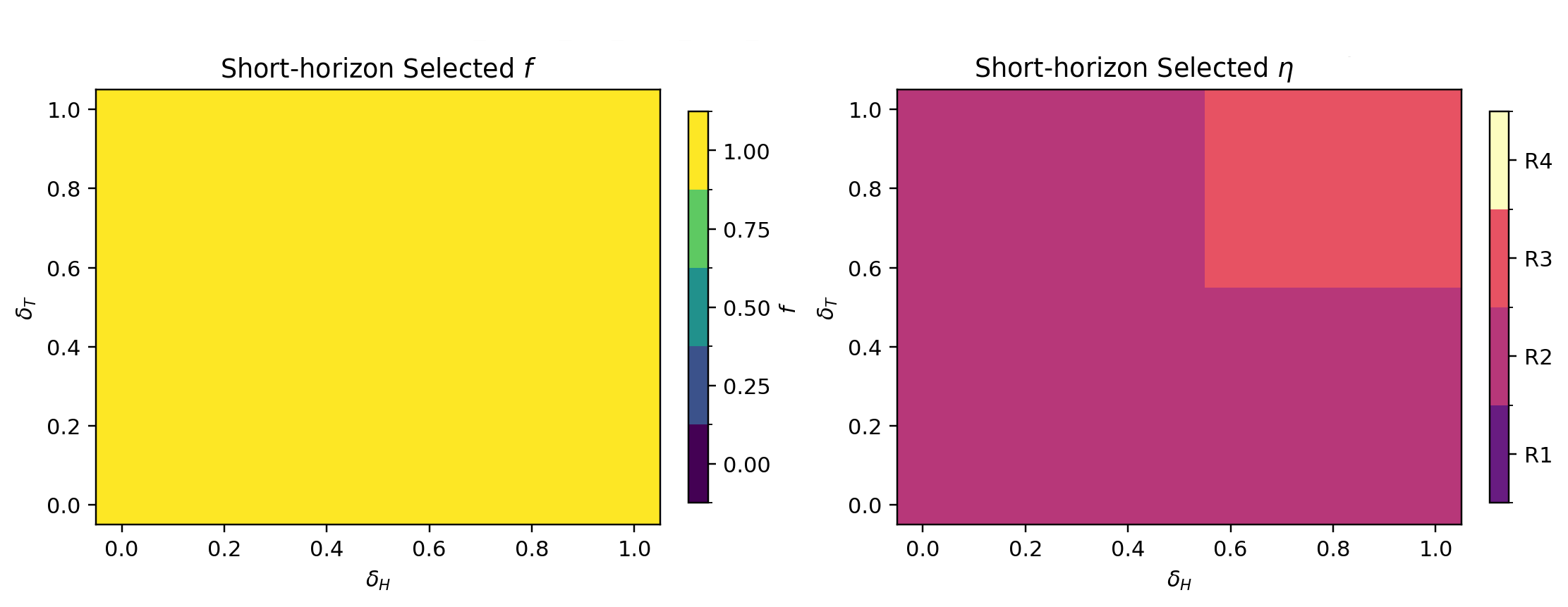}
        \caption{Short-horizon: $(H_t,T_t)=(0.5,0.2)$.}
    \label{fig:12_analytic_policy_high_deficit_low_trust_H0.5_T0.2}
    \end{subfigure}
    \hfill
    \begin{subfigure}[t]{0.48\textwidth}
        \centering
        \includegraphics[width=1\textwidth]{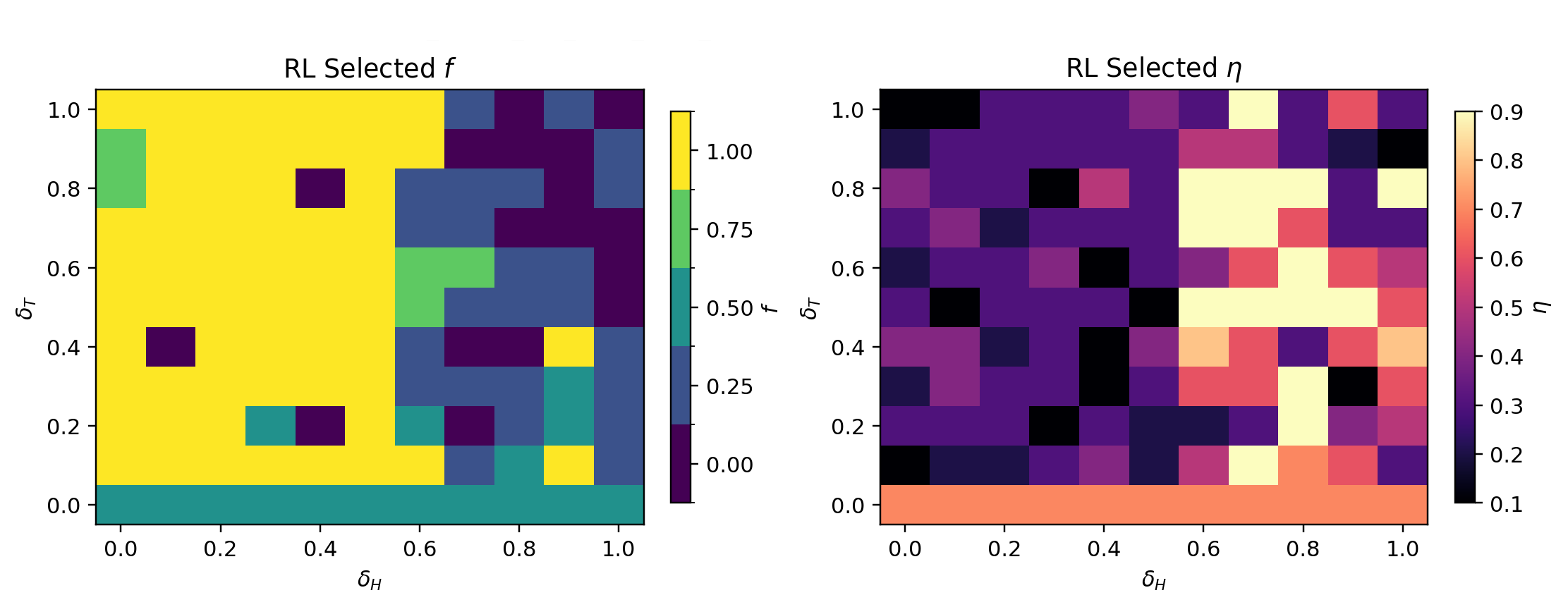}
        \caption{RL: $(H_t,T_t)=(0.5,0.2)$.}
    \label{fig:13_rl_policy_seed0_high_deficit_low_trust_H0.5_T0.2}
    \end{subfigure}
    \hfill
    \begin{subfigure}[t]{0.48\textwidth}
        \centering
        \includegraphics[width=1\textwidth]{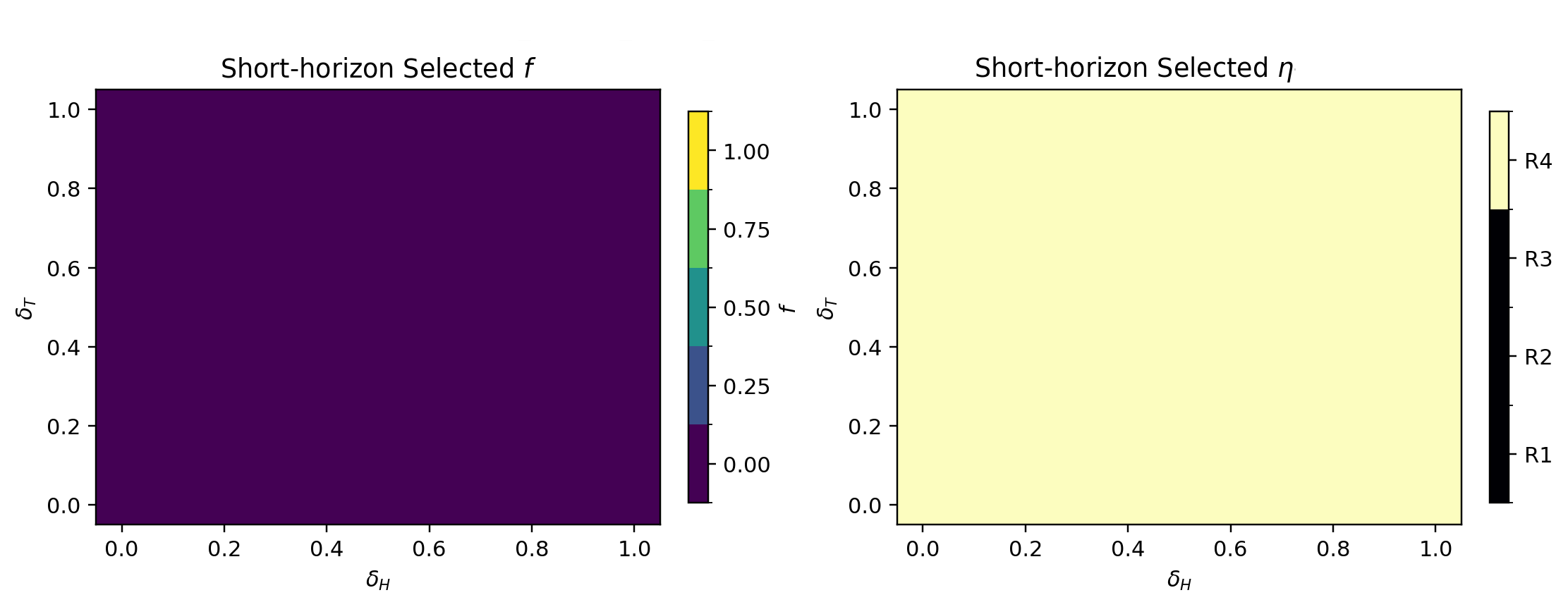}
        \caption{Short-horizon: $(H_t,T_t)=(0.95,0.5)$.}
    \label{fig:14_analytic_policy_near_healthy_H0.95_T0.5}
    \end{subfigure}
    \hfill
    \begin{subfigure}[t]{0.48\textwidth}
        \centering
        \includegraphics[width=1\textwidth]{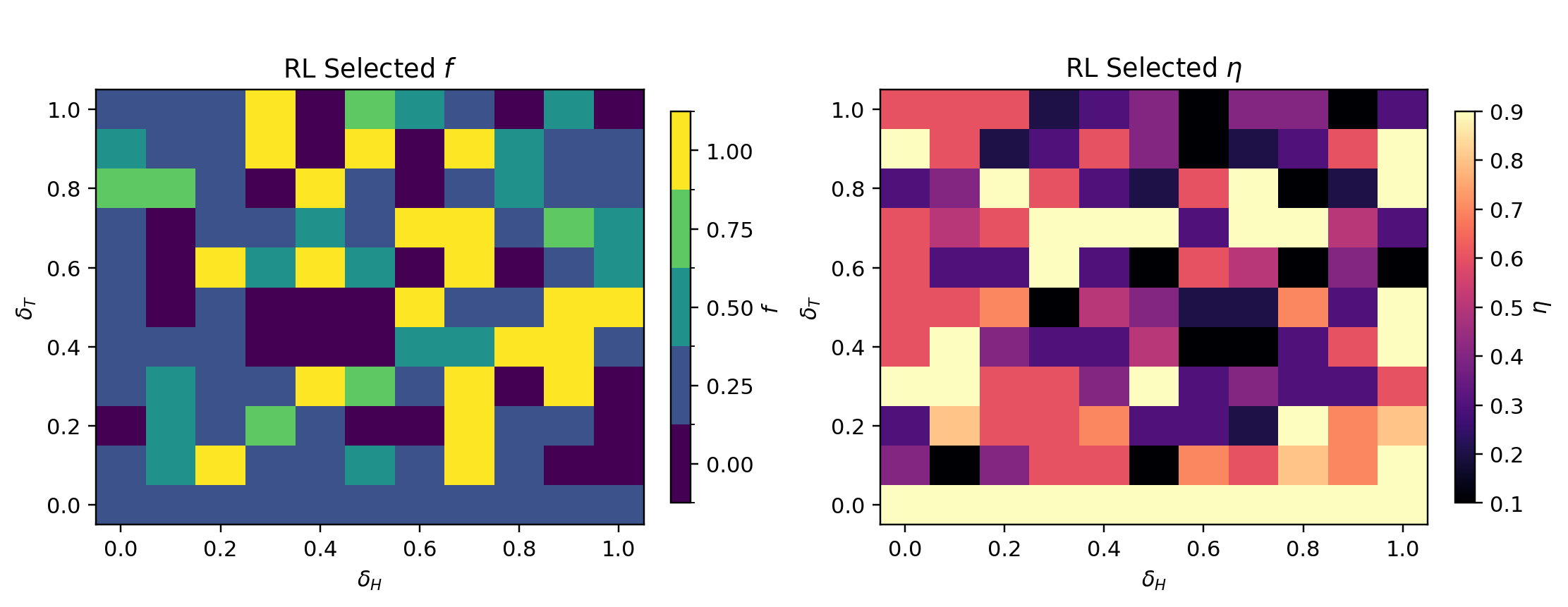}
        \caption{RL: $(H_t,T_t)=(0.95,0.5)$.}
    \label{fig:15_rl_policy_seed0_near_healthy_H0.95_T0.5.png}
    \end{subfigure}
    \caption{Comparing short-horizon policy and RL in terms of frequency and decision threshold $\eta$ for different user types, under three alternative initial health and trust states. All other parameters are unchanged from the main experiments, and the same trained RL (seed 0) is used throughout.}
    \label{fig:initial-state-policy-sensitivity}
    \vspace{-0.12in}
\end{figure*}

\begin{figure*}[!ht]
    \centering
    \begin{subfigure}[t]{0.48\textwidth}
        \centering
        \includegraphics[width=1\textwidth]{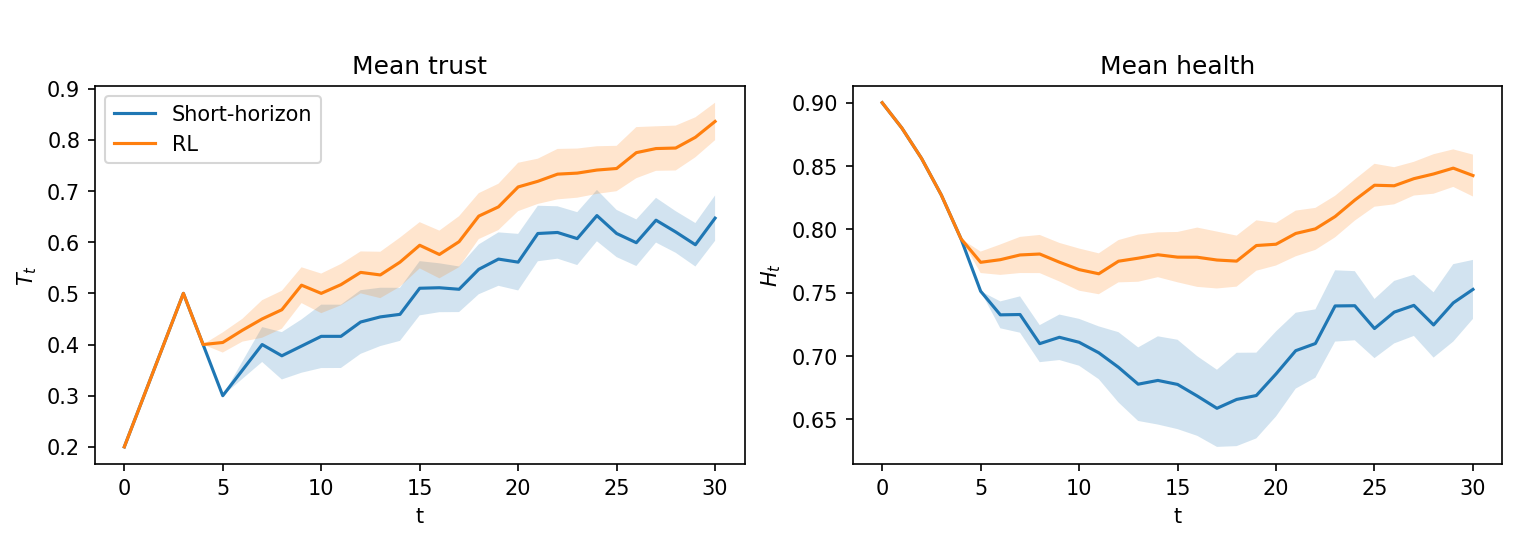}
        \caption{$(H_t,T_t)=(0.9,0.2)$}
        \label{fig:19_seed0_low_deficit_low_trust_H0.9_T0.2_dt0.10_dh0.20}
    \end{subfigure}
    \hfill
    \begin{subfigure}[t]{0.48\textwidth}
        \centering
        \includegraphics[width=1\textwidth]{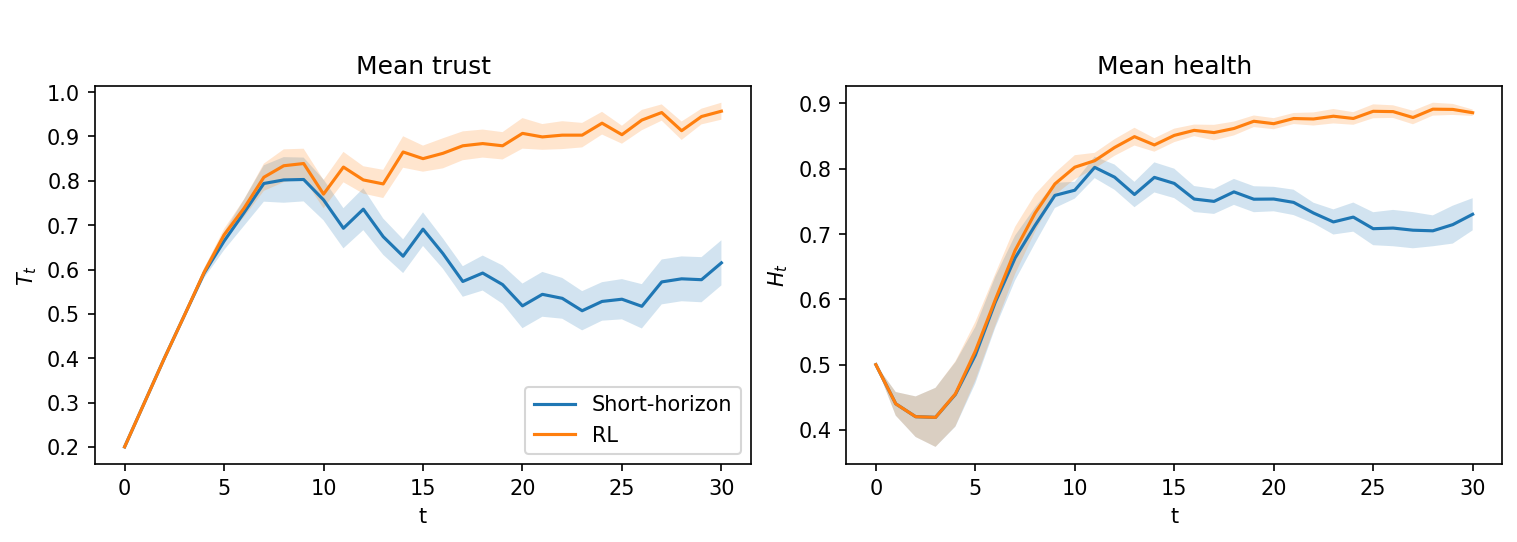}
        \caption{$(H_t,T_t)=(0.5,0.2)$}
        \label{fig:20_seed0_high_deficit_low_trust_H0.5_T0.2_dt0.10_dh0.20}
    \end{subfigure}
    \hfill
    \begin{subfigure}[t]{0.5\textwidth}
        \centering
        \includegraphics[width=1\textwidth]{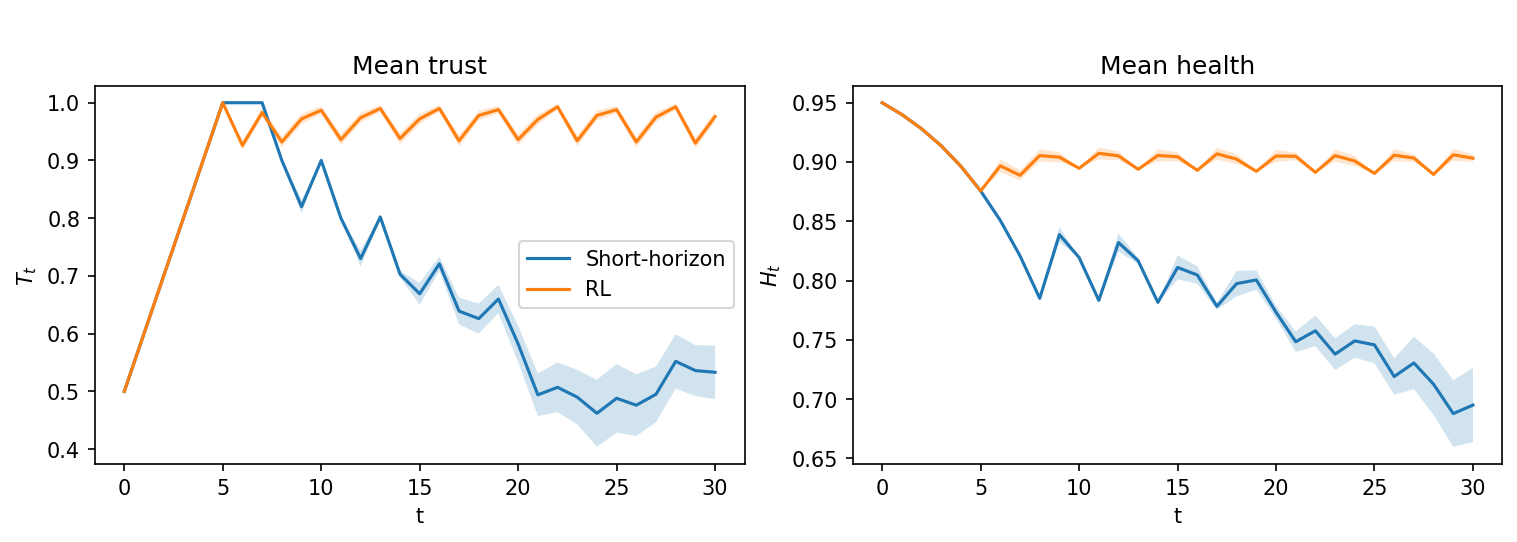}
        \caption{$(H_t,T_t)=(0.95,0.5)$}
        \label{fig:21_seed0_near_healthy_H0.95_T0.5_dt0.10_dh0.20}
    \end{subfigure}
    \caption{Evolution of trust and health under different policies for user type $(\delta_H, \delta_T)=(0.2,0.1)$ under three
    alternative initial states. The same trained RL (seed 0) is used in all three cases.}
    \label{fig:initial-state-trajectories}

\end{figure*}

Figure~\ref{fig:initial-state-policy-sensitivity} first shows how the selected threshold-frequency pairs change across the three initial states. We then study how these changes affect the cumulative reward comparison between the RL and short-horizon policies in Figure~\ref{fig:compare-policies-sensitivity}. Finally, to validate whether the trust building behavior observed in the main experiments persists under different initial conditions, Figure~\ref{fig:initial-state-trajectories} compares the resulting trust and health trajectories.

Figure~\ref{fig:initial-state-policy-sensitivity} shows that the selected policies can change substantially with the initial state. When $(H_t,T_t)=(0.9,0.2)$, the short-horizon policy selects $f=0$ and the highest triggering threshold region for all user types, effectively issuing no recommendations (Figure~\ref{fig:10_analytic_policy_low_deficit_low_trust_H0.9_T0.2}). In contrast, when the initial health is lower while trust is kept fixed, $(H_t,T_t)=(0.5,0.2)$, the short-horizon policy selects the highest recommendation frequency, $f=1$, for all user types, while the selected threshold region varies across user types and generally corresponds to lower triggering thresholds (Figure~\ref{fig:12_analytic_policy_high_deficit_low_trust_H0.5_T0.2}). This change can be understood from the short-horizon tradeoff between health and productivity. When the initial health is relatively high, the immediate health benefit of recommending over the two-step horizon is limited, so avoiding the productivity cost can make no recommendations preferable. When the initial health is substantially lower, the potential short-term health benefit of intervention is higher, making frequent recommendations more attractive despite their productivity cost. For the high health, higher trust initialization $(H_t,T_t)=(0.95,0.5)$, the short-horizon policy again selects the no-recommendation policy throughout (Figure~\ref{fig:14_analytic_policy_near_healthy_H0.95_T0.5}), indicating that the small health degradation leads the policy to favor avoiding the immediate productivity cost.

The RL policy, on the other hand, continues to vary the recommendation frequency across user types under all three initializations. The two high-health initializations,  $(H_t,T_t)=(0.9,0.2)$ and $(H_t,T_t)=(0.95,0.5)$ show broadly similar heterogeneous frequency patterns (Figures~\ref{fig:11_rl_policy_seed0_low_deficit_low_trust_H0.9_T0.2} and~\ref{fig:15_rl_policy_seed0_near_healthy_H0.95_T0.5.png}). When the initial health is substantially lower, $(H_t,T_t)=(0.5,0.2)$, a clearer dependence on $\delta_H$ appears: higher recommendation frequencies are more common for smaller $\delta_H$, while lower and more varied frequencies are selected as $\delta_H$ increases (Figure~\ref{fig:13_rl_policy_seed0_high_deficit_low_trust_H0.5_T0.2}). This greater heterogeneity is consistent with the long-horizon nature of the RL policy, which accounts for how current recommendations affect future trust and compliance differently for each user type, and therefore can favor different actions across user types.

Overall, these results show that the actions selected by both policies depend on the initial state. However, the qualitative distinction observed in the main experiments remains across these alternative initializations: the short-horizon policy can select the same frequency over large portions of user types, whereas the RL policy generally exhibits more variation across user types.

Next, we observe that although the selected policies differ across the initial states, their relative performance remains generally similar. As shown in Figure~\ref{fig:compare-policies-sensitivity}, the RL policy achieves higher cumulative reward for most user types under all three selected initializations, while the short-horizon policy remains competitive or performs better for some user types.

\begin{figure*}[!ht]
    \centering
    \begin{subfigure}[t]{0.32\textwidth}
        \centering
        \includegraphics[width=1\textwidth]{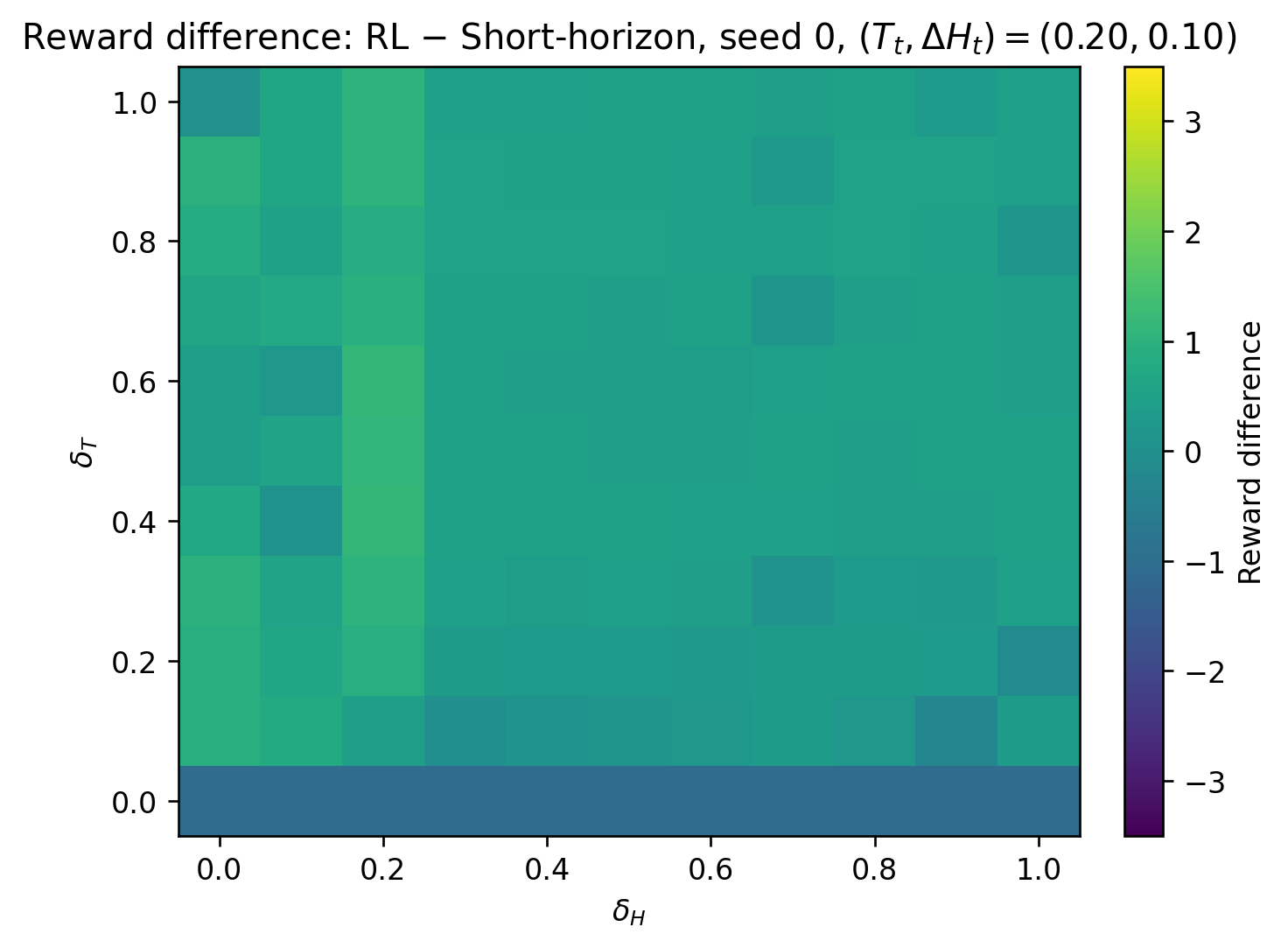}
        \caption{$(H_t,T_t)=(0.9,0.2)$}
        \label{fig:16_fig2_reward_gap_seed0_low_deficit_low_trust_H0.9_T0.2}
    \end{subfigure}
    \hfill
    \begin{subfigure}[t]{0.32\textwidth}
        \centering
        \includegraphics[width=1\textwidth]{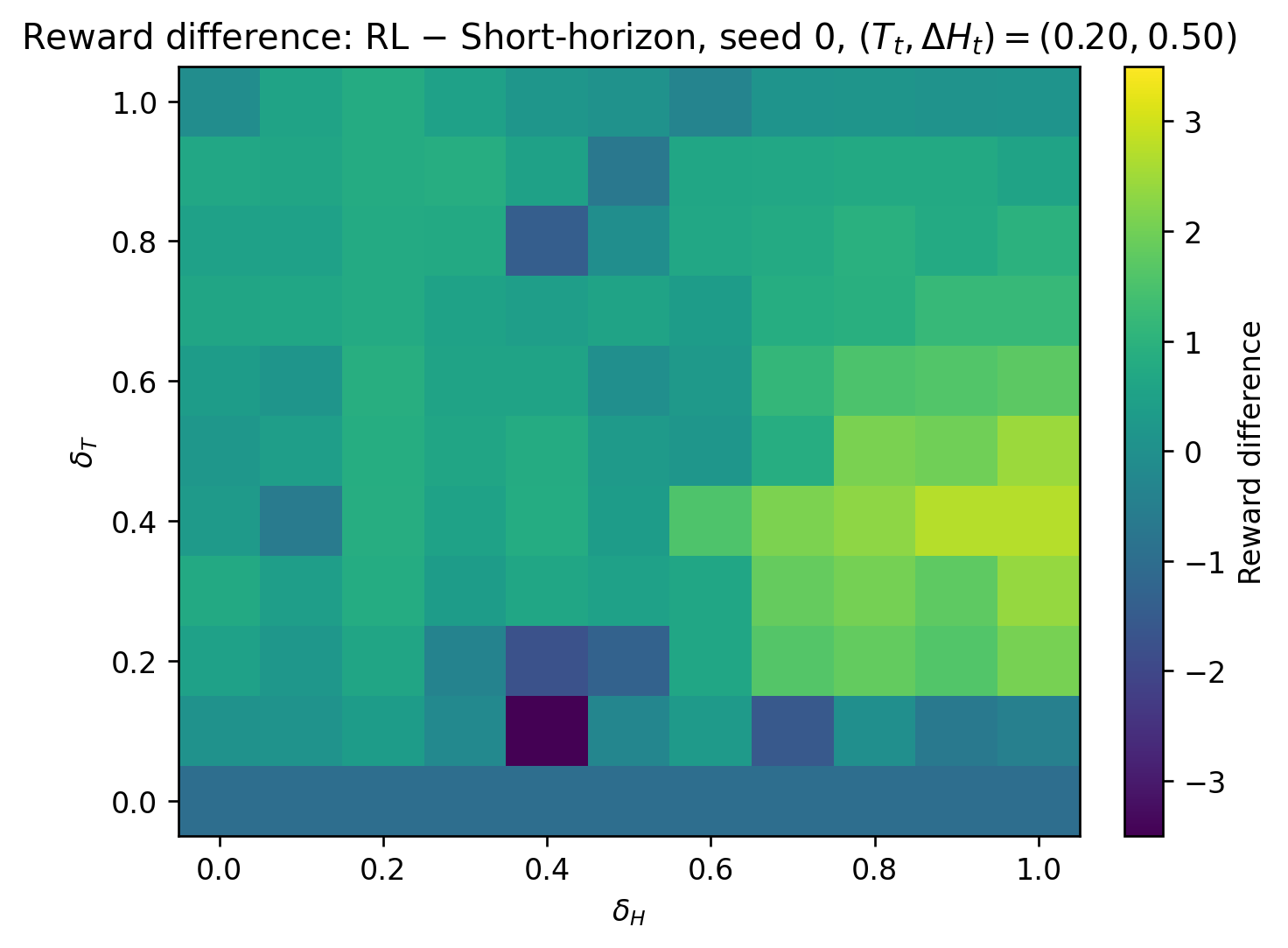}
        \caption{$(H_t,T_t)=(0.5,0.2)$}
        \label{fig:17_fig2_reward_gap_seed0_high_deficit_low_trust_H0.5_T0.2}
    \end{subfigure}
    \hfill
    \begin{subfigure}[t]{0.32\textwidth}
        \centering
        \includegraphics[width=1\textwidth]{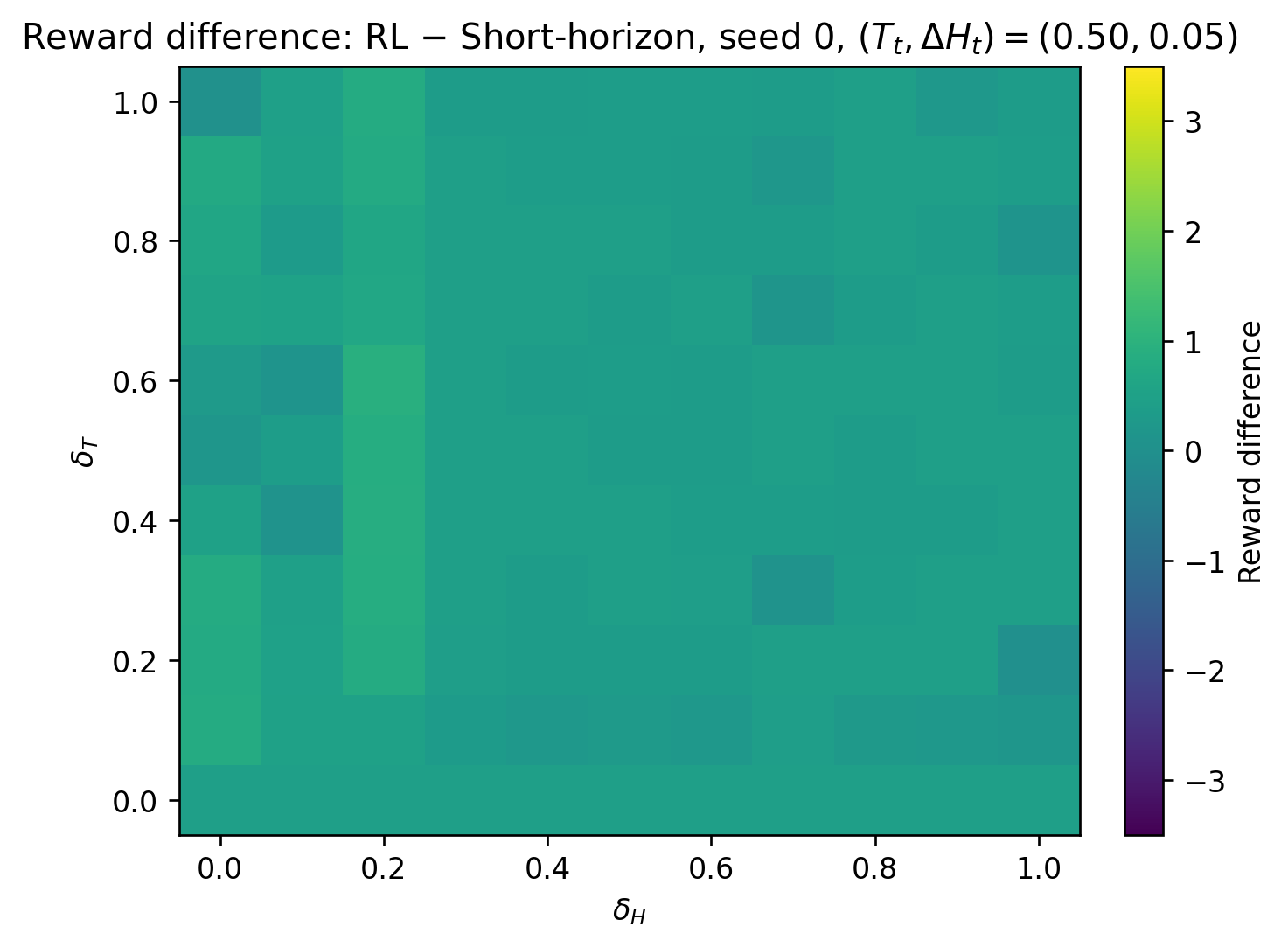}
        \caption{$(H_t,T_t)=(0.95,0.5)$}
        \label{fig:18_fig2_reward_gap_seed0_near_healthy_H0.95_T0.5}
    \end{subfigure}
    \caption{Cumulative reward difference between the RL and short-horizon policies across user types under three alternative initial states. The same trained RL (seed 0) is used in all three cases.}
    \label{fig:compare-policies-sensitivity}
\end{figure*}

\begin{figure*}[!ht]
    \centering
    \begin{subfigure}[t]{0.45\textwidth}
        \centering
        \includegraphics[width=1\textwidth]{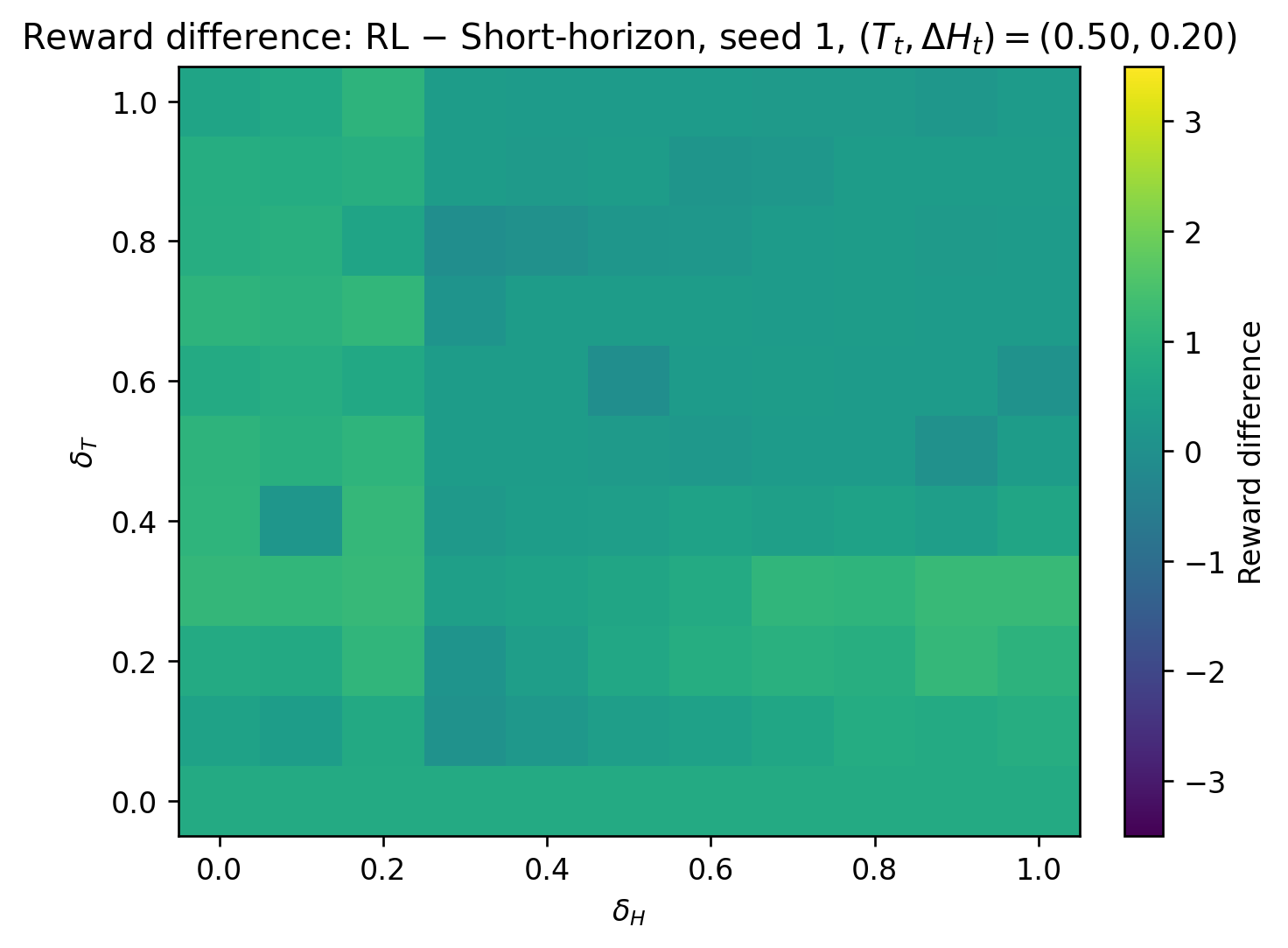}
        \caption{RL training seed 1.}
        \label{fig:22_fig2_reward_gap_seed1_paper_original_H0.8_T0.5}
    \end{subfigure}
    \hfill
    \begin{subfigure}[t]{0.45\textwidth}
        \centering
        \includegraphics[width=1\textwidth]{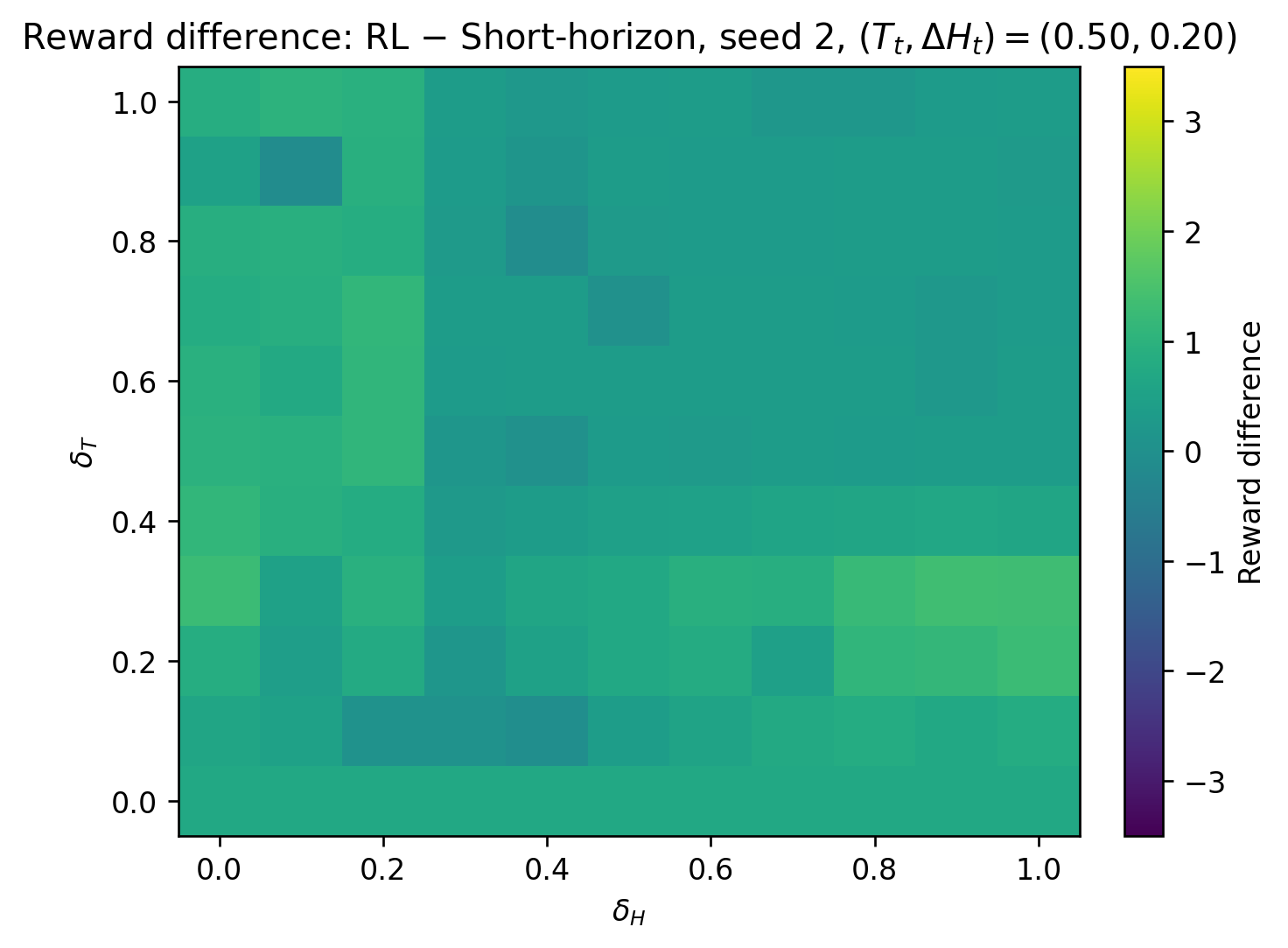}
        \caption{RL training seed 2.}
        \label{fig:23_fig2_reward_gap_seed2_paper_original_H0.8_T0.5}
    \end{subfigure}
    \caption{The cumulative reward difference under short-horizon and RL policies for different user types, for two additional RL training seeds. The
    initial state is fixed at $(H_t,T_t)=(0.8,0.5)$, as in the main
    experiments.}
    \label{fig:compare-policies-reward-seed}
    \vspace{-0.2in}
\end{figure*}

Finally, we examine whether the trust building behavior observed in the main experiments persists under different initial states. We consider the same user type as in
Figure~\ref{fig:4_health-trust_low}, $(\delta_H,\delta_T)=(0.2,0.1)$, while varying the initial state as described at the beginning of this section.

For the high-health, low-trust initialization $(H_t,T_t)=(0.9,0.2)$ (Figure~\ref{fig:19_seed0_low_deficit_low_trust_H0.9_T0.2_dt0.10_dh0.20}), the initial health degradation is $\Delta H_t=0.1<\delta_H$. The short-horizon policy initially avoids recommending, which increases trust, while health gradually deteriorates. As the health degradation becomes larger, the short-horizon policy changes its behavior and begins recommending, allowing health to recover. The RL policy, in contrast, follows a strategy that gradually builds trust over time. Although this is accompanied by an initial health decrease, the policy subsequently recovers and ultimately maintains both higher trust and higher health than the short-horizon policy.

When $(H_t,T_t)=(0.5,0.2)$ (Figure~\ref{fig:20_seed0_high_deficit_low_trust_H0.5_T0.2_dt0.10_dh0.20}), the initial health degradation is already large, $\Delta H_t=0.5>\delta_H$. In this case, both policies initially favor recommendations, leading to improvements in trust and health. As health improves, however, the effect of subsequent recommendations on trust changes. The short-horizon policy responds according to its two-step planning, after which both trust and health begin to decline. The RL policy, by accounting for the longer-term effects of its actions on future trust and compliance, is able to maintain high trust and continue improving health over the longer horizon.

For $(H_t,T_t)=(0.95,0.5)$ (Figure~\ref{fig:21_seed0_near_healthy_H0.95_T0.5_dt0.10_dh0.20}), the initial health degradation is even smaller, $\Delta H_t=0.05<\delta_H$, while the initial trust is higher than in the first case. Both policies initially reach high trust while health decreases moderately. Their trajectories then separate. Under the short-horizon policy, trust and health subsequently decline, since the two-step objective does not account for the further consequences of the current actions. In contrast, the RL policy adjusts its behavior so as to maintain trust at a high level and preserve substantially higher health.

It is also interesting to compare the first and third initializations. Although the short-horizon policy initially selects the same no-recommendation behavior under $(H_t,T_t)=(0.9,0.2)$ and $(H_t,T_t)=(0.95,0.5)$, the resulting long-horizon trajectories are different. In the first case, the short-horizon policy eventually recovers part of the lost health and even improves trust, whereas in the third case both quantities show a continued decline. The RL policy, in contrast, maintains better trust and health trajectories in both cases.

Overall, although the transient behavior depends on the initial state, the main trend observed in the original experiment persists in these experiments: the long-horizon RL policy is better able to maintain trust and use it to improve health over time.

\subsection{Sensitivity to RL Training Seed}

Considering another aspect, to examine whether the previous results are robust to the stochasticity of DQN training, we repeat the training of RL agents using two additional random seeds. We then compare the cumulative reward under short-horizon policy, which is unchanged across seeds, with that under the RL policy (Figure~\ref{fig:compare-policies-reward-seed}). All other parameters are kept the same as in the main experiments, while the RL training seed is changed to 1 in Figure~\ref{fig:22_fig2_reward_gap_seed1_paper_original_H0.8_T0.5} and 2 in Figure~\ref{fig:23_fig2_reward_gap_seed2_paper_original_H0.8_T0.5}. Across both seeds, the performance in terms of cumulative reward remains similar: the long-horizon RL policy achieves higher cumulative reward for most user types. We also continue to observe smaller reward differences for user types with large $\delta_H$ and $\delta_T$, indicating closer performance between the two policies in this region.

\end{document}